\documentclass[journal]{IEEEtran}
\usepackage{cite}
\usepackage{multicol}
\usepackage{lipsum}
\usepackage{fancyhdr}
\usepackage{graphicx,latexsym}
\usepackage{epstopdf}
\usepackage{booktabs}
\usepackage{algorithmic,xcolor}
\usepackage{algorithm,url,epsfig,times, subfigure, setspace, amsmath}
\newcommand{\tc}{\textcolor{black}}
\begin{document}
\title{Multi-path Model and Sensitivity Analysis for Galvanic Coupled Intra-body Communication through Layered Tissue}
%\title{path Characterization for Wireless Galvanic Coupled Intra-body Communication}
\author{Meenupriya Swaminathan, Ferran Simon Cabrera, Joan Sebastia Pujol, Ufuk Muncuk,\\ Gunar Schirner,~\IEEEmembership{Member,~IEEE,} and~Kaushik R. Chowdhury,~\IEEEmembership{Member,~IEEE}  
\thanks{Meenupriya Swaminathan, Ufuk Muncuk, Gunar Schirner, and Kaushik R. Chowdhury are with the Electrical and Computer Engineering Department, Northeastern University, Boston, MA 02115 USA. email:\{meenu,muncuk.u,schirner,krc\}@ece.neu.edu. Ferran Simon Cabrera and Joan Sebastia Pujol are with Department of Electronic Engineering, Universitat Polit\`{e}cnica de Catalunya, Barcelona, Spain Email:ferran@nomis.es. This work was performed when the authors were visiting researchers at Northeastern University.}}
\maketitle
\IEEEpeerreviewmaketitle
\pagestyle{plain}
\thispagestyle{fancy}
\pagestyle{fancy}
 \setcounter{page}{1}
 \pagenumbering{arabic}
 \begin{abstract}
New medical procedures promise continuous patient monitoring and drug delivery through implanted sensors and actuators. When over the air wireless radio frequency (OTA-RF) links are used for intra-body implant communication, the network incurs heavy energy costs owing to absorption within the human tissue. With this motivation, we explore an alternate form of intra-body communication that relies on weak electrical signals, instead of OTA-RF. To demonstrate the feasibility of this new paradigm for enabling communication between sensors and actuators embedded within the tissue, or placed on the surface of the skin, we develop a rigorous analytical model based on galvanic coupling of low energy signals. The main contributions in this paper are: (i) developing a suite of analytical expressions for modeling the resulting communication channel for weak electrical signals in a three dimensional multi-layered tissue structure, (ii) validating and verifying the model through extensive finite element simulations, published measurements in existing literature, and experiments conducted with porcine tissue, (iii) designing the communication framework with safety considerations, and analyzing the influence of different network and hardware parameters such as transmission frequency and electrode placements. Our results reveal a close agreement between theory, simulation, literature and experimental findings, pointing to the suitability of the model for quick and accurate channel characterization and parameter estimation for networked and implanted sensors.
\end{abstract}
\noindent\begin{keywords}
Intra-body communication, galvanic coupling, channel model, circuit model, implanted sensors/actuators, tissue safety.
\end{keywords}
\section{Introduction}
Intra-body networks (IBNs) promise to usher in dramatic improvements in personalized medicine, implant-based in-situ monitoring, controlled drug delivery, and activity based muscular/neuro stimulation, among others. In this paradigm,  micro-scale sensors and embedded actuators may communicate with each other for automatic, real time response, or the sensors transmit wirelessly to a remote monitoring entity that aggregates and monitors the signals generated within the body. Moreover, the sensors may themselves be programmed with new instructions over time, such as activating specific bio-marker receptors for various patient conditions and medical check points. This closed loop system makes it  possible for continuous monitoring without invasive techniques, reduces the delay and human-error in processing the data. As an example case study, diabetic patients frequently self-monitor blood glucose concentrations using small blood samples obtained through a finger prick, and then administer multiple injections of insulin each day or use an insulin pump. However, the insulin is often slow reacting, leading to the possibility of overdose, and the glucose level is only checked at specific intervals, such as after meals. We envisage that the IBN composed of implanted plasma glucose sensors, aided by our implant communication, will continuously sample the accurate glucose level and transmit the data to an embedded insulin pump. The latter will project the patient's glucose level based on current level and past history, and release trickle amounts of insulin, all without human intervention. In addition, specialists can also study the response of the person to the specific insulin type, program any adjustment in dosage, or alter the sensing duty cycle. This same scenario for RF based interstitial fluid glucose sensor, becomes resource heavy and environmental dependent extending to atleast six feet around the body.  

\begin{figure}[b]
 \centering
\includegraphics[width=8cm]{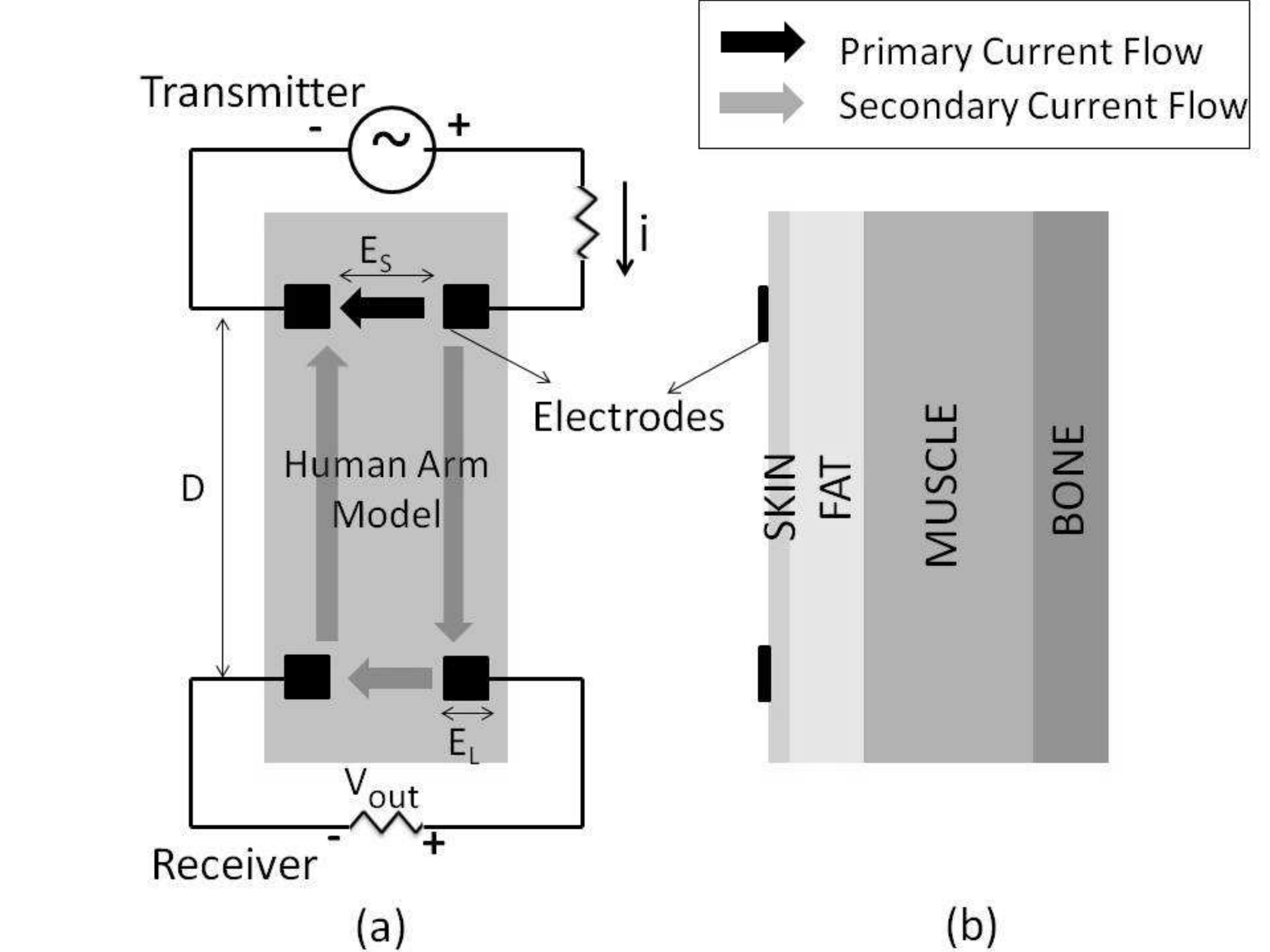}
 \caption{ \label{fig:GC} Galvanic coupling setup on skin surface with multiple tissue layers (a) Top view  (b) Lateral view}
 \end{figure}
 
For IBNs, the retrieval of the sensors for battery replacement becomes impractical, requiring efforts on reducing energy consumption for data aggregation and communication. We shall demonstrate subsequently in this paper that over the air radio frequency (OTA-RF), inductive and ultrasonic \cite{miandus} form of intrabody communication consumes energy at a higher scale, as compared to our proposed approach. Moreover, our choice of using \textit{galvanic coupling} as method for transmitting electrical signals is also motivated by the high water content within the human body, which facilitates the propagation of low frequency waves. While OTA-RF communication is well understood despite its high absorption level within the body, a unified analytical model for the channel gain for weak electrical signal propagation through various tissue layers remains in a nascent stage. The key contribution in this paper is formulating closed form channel gain expressions for IBN by first building a three dimensional multi-layered tissue equivalent circuit model. Our analysis allows reproducibility of results, and is able to accurately predict the channel gain across the skin as well as across and through the inner body tissues. It can accommodate a variety of transmitter-receiver distances, electrode separations and dimensions, various depths of implant embeddings, choice of operating frequency, and tissue thicknesses. 
 
\subsection{Wireless communication through galvanic coupling:} 
In galvanic coupled communication, a pair of electrodes within a given IBN node couple a weak electric signal of around $1 \mathrm{mW}$ to the body tissue \cite{Gal1}, which is first modulated by the sensor data. The induced field in the tissue is well below the permissible limit \cite{safety,ICNIRP}, and additional design considerations are further discussed in section \ref{sec:safety}. Majority of the induced current that is coupled to the body passes through the return path of transmitter (represented by black arrow in Fig.\ref{fig:GC}) and a minor part (illustrated by gray arrows) propagates through the body. The difference in potential is detected by the electrode pair of a receiver node. The receiver demodulates the signal to receive the sensor data. Note that there is no common ground required here, as in the case of capacitive coupling \cite{capa1}. A characteristic feature of galvanic coupled communication is that the signal has a dominant component propagating through the inner tissue layers, even when the transmitter is placed on the surface \cite{piganisotropic,ICNIRP}. Thus, apart from being more energy efficient compared to OTA-RF, the IBN communication also becomes less impacted by environmental noise. A carefully designed coupling apparatus with an optimized signal amplitude and frequency gives rise to a dominant signal component that can be guided to traverse through specific part of the body. Thus, multiple concurrent transmissions along the same body becomes possible, leading to new challenges in interference-free operation. This behavior differs from OTA-RF propagation, wherein other transceivers must be silenced owing to the broadcast nature of the medium. 
 
\subsection{Research motivation:}
For establishing communication links among the IBN nodes, the tissue channel needs to be analyzed for selecting optimal propagation characteristics in order to safely and reliably transfer information. Our work on an analytic model for building a reliable human tissue communication channel is motivated by the fact that in-vivo tissue experiments are not always possible, commercially available phantoms do not accurately reflect the tissue heterogeneity, and electrical propagation characteristics over a wide frequency range. Human body is composed of multi-layered tissues each with its own signal propagation characteristics. Tissue impedance calculations should include this multi-layer phenomenon for accurate channel estimations. \tc{The state of the art has been mainly restricted to a single tissue communication such as on-surface (i.e., with the transmitter and receiver placed on the skin), with a limited investigation in muscle~\cite{musclephantom}, that analyzes only three directions of current flow. Our model completely changes this analysis using practical assumptions of the tissue electrical properties, where four directions of current flow (the additional direction involving current passing into lower/upper tissue layers) is possible. To the best of our knowledge, this comprehensive treatment of galvanic coupling-based channel model has not been derived before, and for successful communication between implanted sensors, it is essential for characterizing the transverse path from one tissue to another.}

Moreover, for a detailed analysis on the implant data link through tissues, the communication channels along tissues needs to be characterized individually as skin to skin (S-S), skin to muscle (S-M), muscle to skin (M-S) and muscle to muscle (M-M) paths, among others. The field distribution arising out of the galvanic coupled multi-layered inner tissue that includes the above mentioned intra-body scenarios needs further investigation, as no reproducible analytic model exists that has been verified through experiments. %The accuracy of models that simplifies the complex structure of human body, such as using a concentric cylinders to represent the forearm is to be verified highlighting the importance to validate the results from modeling studies with experimental measurements and clinical trial data available.

We summarize the main contributions of our work as follows:
\begin{itemize}
\item We derive a three dimensional multi-layered human forearm Tissue Equivalent Circuit model (TEC) for analyzing the communication channel through the surface and inner tissue-layers. \tc{Our reproducible expression involves a large number of configurable parameters (over 10), which can comprehensively capture the various design intricacies of GC-IBN-based communication.}
\item Our theoretical approach is validated with previously conducted experiments for on-skin communication. \tc{Interestingly, our model indicates a tighter match with previously obtained measurements, than what was possible using existing models.} We also include additional validations through measurement studies conducted on porcine tissue.
\item For verifying the accuracy of the multi-tissue analysis, we construct a 3D model of the human forearm using finite element simulation. The simulator  captures minute aspects of the signal propagation through the inner tissues. This allows the simulation to be used for quick analysis of future network designs for situations where intra-body testing is not immediately feasible.
\item We ensure that safety considerations are incorporated based on electric current distributions inside tissues, and we identify the ideal transmission frequency ranges that provide the best performance. 
\item We analyze the model for various parameters like tissue thickness and electrode dimensions/separations and provide insights on suitable implant positions inside the tissues.
\end{itemize}

The rest of the paper is organized as follows: Section~\ref{sec:related} gives the related work. We formulate our analytical model based on a circuit equivalent construction for the human forearm in Section~\ref{Sec:Model} with the corresponding simulation model and safe signal generation conditions described in Section~\ref{FEM}. The model verification and analysis of the model parameters are given in Sections~\ref{Results} and~\ref{Sec:sense}, respectively. Measurements based on porcine tissues are presented in Section~\ref{Sec:pig}, and finally, Section~\ref{Conclusion} concludes the paper. 

\section{Related Work}
\label{sec:related}
Among the different techniques available for modeling the tissue electrical behavior, quasi-static approximations, \cite{QS1,QS5}, full wave numerical techniques such as Finite Difference Time Domain Method (FDTD), Finite Element Analysis (FEA) \cite{FEM1,FEM2}, and Equivalent Circuit Analysis (ECA) based modeling are the main approaches. The quasi-static field distribution analyses are computationally efficient. However, they only represent low frequency approximations to Maxwell's equations and cannot be relied on for high frequency applications. Field analysis using FDTD and FEA are flexible and accurate but require a great deal of time for computing, and find limited application in a  rapid  deployment of an IBN. The ECA model offers a simple transfer function valid for a wide range of frequency, with the advantage of accurate and instantaneous gain computation making is feasible for IBN deployment for time-sensitive healthcare applications. However, most of the existing approaches \cite{distri,lumped2,lumped4} consider single tissue layer with limited flexibility, which we aim to overcome in our proposed work. Additionally, works that consider the multi-layer effect \cite{lumped4} include only bidirectional signal propagation paths (longitudinal and cross paths) between transmitter and receiver. The direct path between the transmitter terminals that  depends on the underlying tissue impedance is assumed to be measurable at the electrode attachment site \cite{FEM2,Gal8}, which limits its practicality. Also, the transverse path from one tissue to other that depends on the tissue thickness is neglected. The tissue equivalent model needs to be asymmetric as opposed to the existing models to account for dissimilar dimensions, tissue heterogeneity, and non-identical electrode set-up at transmitter and receiver, which significantly complicates the analysis. 

\section{Three Dimensional Tissue Equivalent Circuit Model of Human Forearm} \label{ECM}
We aim to build a Tissue Equivalent Circuit (TEC) model that should quickly provide an estimate of the channel gain based on the choices of input frequency, transmitter-receiver locations, distance and separation between their electrodes. Our model uses some easily obtained physiological factors, such as dimensions and hydration levels. We specifically design the model for the human forearm, with the individual tissue impedance obtained from their electrical properties. The corresponding dimensions are average values for an adult male. We derive this model next using the frequency dependent electrical properties of tissues. 

\subsection{Tissue Impedance:}
Living tissue is composed of both movable charges and movement restricted dipoles. Hence, it can be characterized as an imperfect dielectric medium. When an array of electricity conducting cells are excited by an external electrical signal, each cell activates its neighbor, enabling signal propagation through different paths dictated by the cell structure and the frequency of operation. Low frequency signals cannot penetrate the high impedance cell membrane, and so it takes the circuitous path through extra-cellular fluid. As opposed to this, high frequency signals pass through intra-cellular fluid by penetrating the cell membrane. Thus, the cell membrane gives a capacitance effect, allowing the passage of only high frequency components. 

\begin{figure*}[t]
 \centering
 \includegraphics[width=18cm, height=7.5cm]{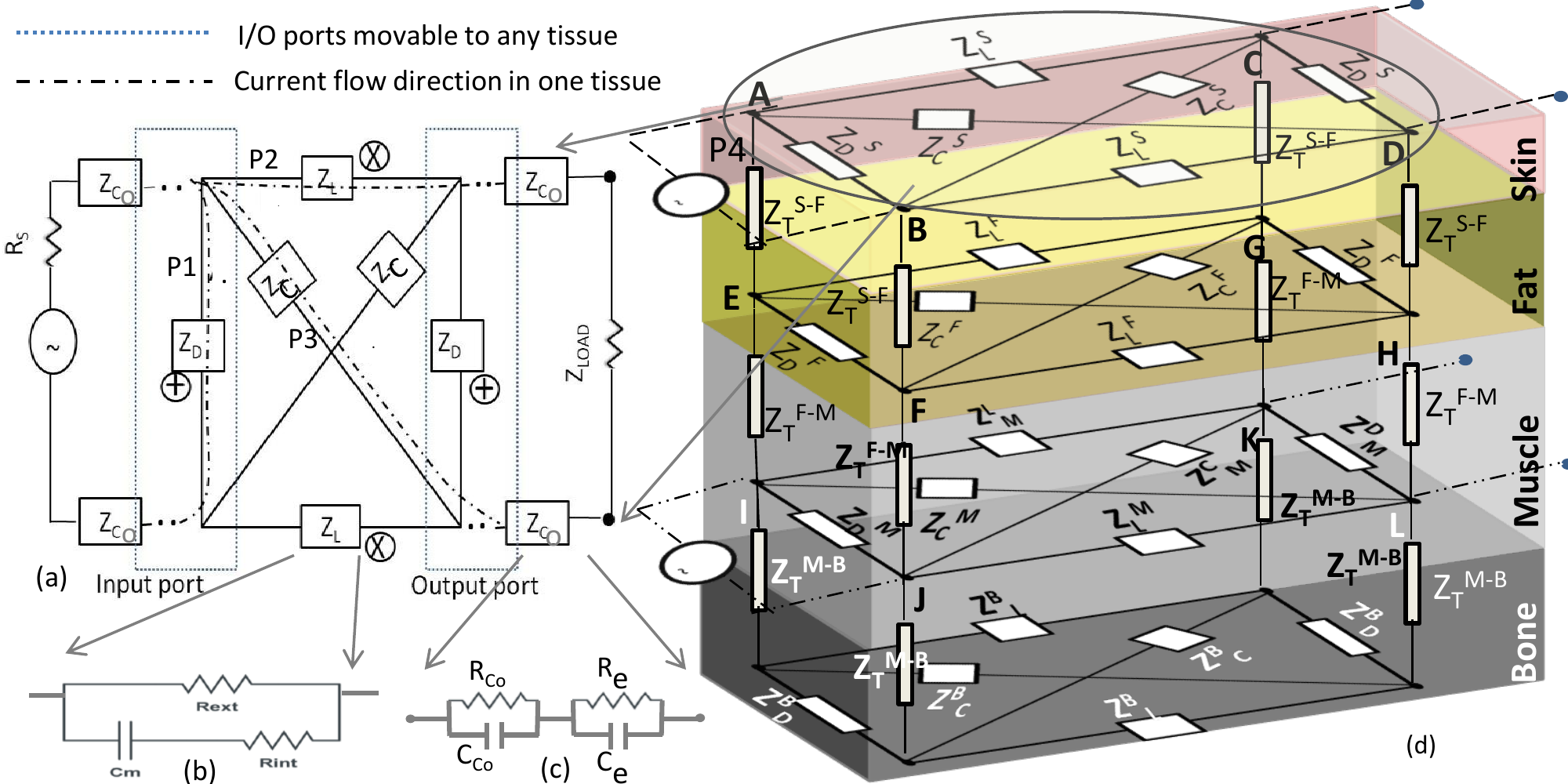}
 \caption{(a) Equivalent circuit of  a single tissue layer, (b) Equivalent circuit of single biological cell, (c) Equivalent circuit of electrode and coupling impedance, and (d) 3D Circuit Model for the forearm as a layered dielectric}
 \label{fig:3d}
 \end{figure*}

Under $100\,\mathrm{MHz}$, the dimensions of human body and implants are small compared to the signal wavelength, and hence, we undertake the analysis using {\em lumped elements}. Using the frequency dependent electrical properties of live tissues (conductivity $(\sigma)$ and permittivity $(\epsilon)$), a simple biological cell can be modeled with Resistance $R_{ext}$, $R_{int}$  (representing dissipation loss), and a capacitor $C_{m}$ (representing the charge holding ability), connected as shown in Fig.\ref{fig:3d}(b). We use the approach in \cite{prop3} to derive the electrical properties of human tissues as given below.
\begin{equation}
\epsilon = \epsilon_0 \epsilon_r = \epsilon_0( \epsilon_r'-j(\epsilon''+\frac{\sigma}{\omega \epsilon_0}))
\end{equation}
where $\epsilon'$ is the dielectric constant and $\epsilon''$ is the out of phase loss factor, expressed in terms of complex permittivity $(\epsilon^*)$ as, 

\begin{equation} \label{eps_eqn}
\epsilon^*= \epsilon'  -j \epsilon''
\end{equation}
\begin{equation} \label{epssingledash}
\epsilon' = \epsilon_\infty + \frac{\epsilon_s - \epsilon_\infty }{1+\omega^2\tau^2}
\end{equation}
\begin{equation} \label{epsdoubledash}
\epsilon'' = \frac{(\epsilon_s - \epsilon_\infty)\omega\tau }{1+\omega^2\tau^2}
\end{equation}
In the above set of equations, $\epsilon_\infty$ and $\epsilon_s$ are dielectric constants at very high and very low frequencies, $\omega$ is the angular frequency measured as $2\pi \times \mathrm{frequency}$ and $\tau$ is the dielectric relaxation time given by $X/R$, where X is the reactive component from capacitance effect. 
%The dielectric loss tangent $(tan\  \delta)$ represents inherent dissipation of electromagnetic energy given by $\left( \displaystyle \frac{\omega \epsilon'' + \sigma}{\omega \epsilon'}\right)$. 

Using (\ref{epssingledash}) and (\ref{epsdoubledash}),  the tissue admittance using RC elements can be calculated as,
\begin{small}
\begin{equation} \label{eqn:imp}
Y = G_{ext}+\frac{1}{R_{int}+j X_C} = F_W \left(\sigma M_1 + \frac{1}{ \sigma \kappa M_1 +j \omega \epsilon M_2} \right)
\end{equation}
\end{small}
where Z is the impedance, G is the conductance, $M_1$ is the ratio of cross sectional area (A) and length of the channel (L) decided by the direction of impedance measurement and  while $M_2$ is the ratio of A and thickness of channel as explained in section \ref{Sec:Model}, $F_W \in [1,10]$ is the correction factor accounting for variation in dielectric properties with respect to tissue water content water distributions \cite{fw} that can be determined using non-invasive hydration testing and $\kappa$ is the ratio of external to internal cell resistance. %chosen to be $0.7$. 
We assume that the other tissue properties can be estimated without actual measurement such as tissue thickness approximation using body mass index (BMI), bio-electrical impedance analysis or triceps skin fold thickness.   
\label{2-port}
 
\subsection{Single Tissue Equivalent Model} 
Prior to the complete modeling of the forearm, the equivalent circuit of a single galvanic coupled tissue is calculated using four impedance. These impedance values are derived as follows, based on the four paths taken by an injected current. These are marked as P1, P2 and P3 in Fig.\ref{fig:3d}(a) for a single tissue layer, and the fourth path from one tissue layer to a neighbor is shown as P4 in Fig.\ref{fig:3d}(d).
\begin{itemize} 
\item Path P1 is the primary return path offering the direct impedance $Z_D$ that channels the majority of current from the terminal to ground electrodes in the transmitter. In this case the factor $M_1$ given in (\ref{eqn:imp}) takes the form $  (E_L\  \times \  T)/E_S$, where $E_L$ is a side of the square electrode, T is tissue thickness and $E_S$ is the terminal-reference electrodes separation distance in transmitter and in receiver that are assumed to be the same if not specified. To distinguish them if they are different, we use the representation $E_{ST}$ for the transmitter electrode separation and $E_{SR}$ for the receiver electrode separation. 

\item Path P2 serves as a pathway for a small portion of current directed towards the receiver electrodes through longitudinal impedance $Z_L$, between the transmitter and receiver electrodes. $M_1$ of $Z_L$ is calculated as $ (E_L \times \  T)/D$, where, $D$ is the transmitter-receiver separation distance. 

\item Path P3 is the electric current propagation path from source terminal in transmitter to the reference terminal in receiver through cross impedance $Z_C$. $M_1$ in this case becomes $(\sqrt{2}E_L\  \times \  T)/(\sqrt{D^2+E_{ST}^2})$. In all the above cases, $M_2$ is chosen to be the tissue thickness. 

\item Path P4 is the electric current propagation path to adjacent tissue layer through transverse impedance $Z_T$. To compute this impedance, $M_1$ is substituted with $T/A_{e}$, where, $A_{e}$ is the electrode area. In this case, $E_S$ becomes the channel thickness.
\end{itemize}

We also include the effect of the coupling impedance offered by the contact between the electrode and the tissue interface in the derivation of channel characteristics, as it determines the amount of signal entering into the tissue. This impedance denoted as $Z_{Co}$ (refer Fig.\ref{fig:3d}(a)), is calculated next.

\paragraph{Electrode-Tissue Coupling Impedance}
 The coupling impedance is a function of frequency, area of contact, tissue hydration, electrode material and surface treatment. To calculate the equivalent impedance at the electrode-tissue interface, we follow the approach in \cite{bodyimpedance}, where the interface is modeled as shown in Fig.\ref{fig:electrode}(b). Here, 
\begin{equation}
Re = K_{1}f^{m} /A_e 
\end{equation} 
and 
\begin{equation}
Xe= 1/wC_{e}= K_{2}f^{m'} /A_e,
\end{equation}
where, f is the frequency of operation, $K_1$ depends on the electrode material. $K_2$ lies within the range (0,1) based on the tissue hydration and surface treatment, m and m' are constants for diffusion control and for activation control.  %$K1 = 5.6x10^3; k2= 3.2x10^4 (\Omega cm^2)$ (copper).
The dots in Fig.\ref{fig:3d}(a) represents the possibility of attaching $Z_{Co}$ to any tissue based in the channel under study. \tc{For instance, along the S-M path, the coupling impedance, $Z_{Co}$, at the transmitter and receiver positions are included in the direct impedance $Z_D$ at each position. $Z_D$ at the transmitter side is represented as $Z_{DT}$, and that corresponding to the receiver side of the muscle is represented as $Z_{DR}$}. 

For developing a tractable model, we assume uniform transverse tissue thickness along the paths indicated by $\bigoplus$ in Fig.\ref{fig:3d}(a). However, it is possible to introduce asymmetry in the model by varying the electrodes separation $E_S$, $E_D$ and/or T at transmitter and receiver as analyzed in Section.\ref{Sec:sense}. Anisotropism can also be introduced into the model by assuming that the transverse impedance is larger than the longitudinal impedance \cite{aniso}. 
\subsection{Modeling for Forearm}\label{Sec:Model}
\begin{figure}[t]
 \centering
 \includegraphics[width=8cm,height=4.5cm]{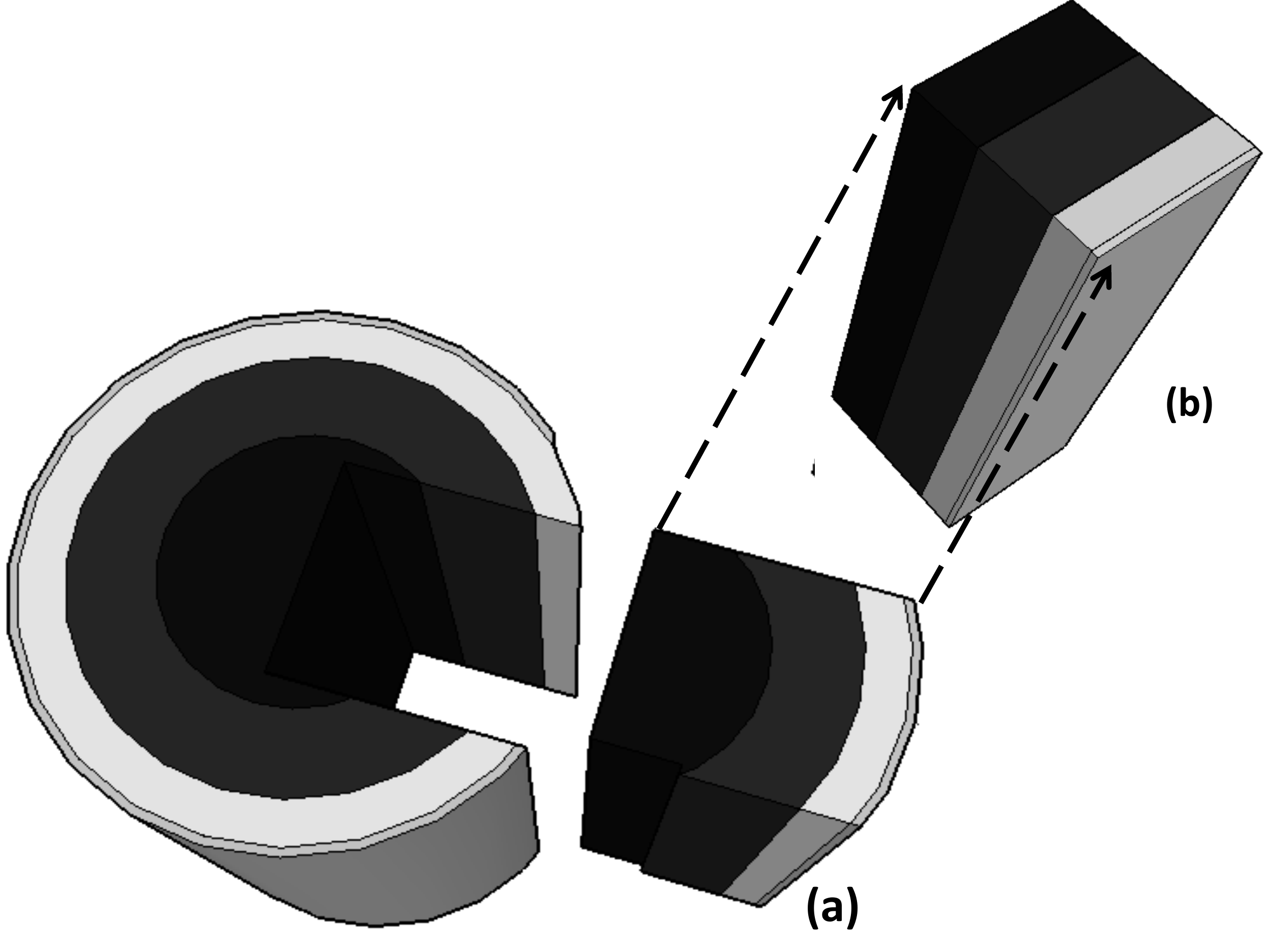}
 \vspace{-4mm}
 \caption{Rectangular layered approximation of longitudinal section of human arm (a) Section of cylindrical arm (b) Cubical approximation}
 \label{fig:armcut}
 \end{figure}

We approximate a longitudinal section of galvanic coupled human forearm (refer Fig.\ref{fig:armcut}) as multi-layered dielectric block with four tissue layers - outer dry skin, fat, muscle and cortical bone (hard outer covering of bone) layers of thickness 1 mm, 7 mm, 15 mm and 20 mm respectively. The parameters such as T, D, $E_S$, and $E_L$ are added as variables in the impedance calculation. The benefit of this equivalent circuit analysis modeling approach is that it uses a simple first-approximation for the voltages and currents that are likely to be observed at different points within the given tissue layer during signal propagation. \tc{The rectangular model (Fig.\ref{fig:armcut}.(b)) enables direct and easier computation of impedances in individual directions. Moreover, it can be extended to any part of the body, such as thorax}.

In the following multi-layer discussion, the superscript $i$ and $j$ denote a specific tissue layer, i.e., $i, j \in \{S, F, M, B\}$, with the substitutions of $S$ for skin,  $F$ for fat, $M$ for muscle, and $B$ for bone. The single tissue impedance $Z_D$ and $Z_L$ in Fig.\ref{fig:3d}(a) become $Z_D^{i}$, $Z_L^{i}$, $Z_C^{i}$ and $Z_T$ takes the form $Z_T^{i-j}$, denoting path from layer i to j. The circuit in Fig.\ref{fig:3d}(d) is used to model the flow of current through skin, fat, muscle and bone in the forearm. 
The S-S path characteristics are studied with the transmitter electrodes (across nodes A and B) and receiver electrodes (across nodes C and D) both coupled on the skin surface (depicted dashed lines in Fig.\ref{fig:3d}(d)). The transmitter and receiver are moved to the muscle tissue for analyzing the M-M path (shown as dot-dot-dash lines). The transmitter is coupled to the skin and receiver is moved to the muscle for the S-M path and vice-verse for the M-S path. 

The circuit shown in Fig.\ref{fig:3d}(d) has four tissue layers with 20 tensions (including the terminal branches) and 16 equations and is solved Kirchhoff's Current Law (KCL). The four complex admittance values of each tissue are calculated using (\ref{eqn:imp}). The node with the source terminal attachment becomes the starting node and reference terminal of the transmitter is chosen as the reference node. The current equation for the first node A based on the difference in node voltage is given below. 
\begin{equation}
\frac{V_{A}-V_{B}}{Z_D^S}+\frac{V_{A}-V_{C}}{Z_L^S}+\frac{V_{A}-V_D}{Z_C^S}+\frac{V_{A}-V_{E}}{Z_T^{S-F}}= I
\end{equation}
where $V_X$ is the voltage estimated in node X, $\forall X \in \{A,B,C,..\}$, I is the input current given by $V_{IN}/Z_{in}$ and $Z_{in}$ is the input impedance across transmitter terminals. Similarly, using the following equations, the voltage difference detected on skin for S-S path can be solved across the nodes C and D.
\begin{equation}
\frac{V_{C}-V_{A}}{Z_{L}^S}=\frac{V_{C}-V_{B}}{Z_C^S}+\frac{V_{C}-V_{D}}{Z_{DR}^S}+\frac{V_{C}-V_{G}}{Z_T^{S-F}}
%%V_C = Z_L \left( \frac{V_C - V_B}{Z_C^S}+\frac{V_C - V_D}{Z_{DR}^S}+\frac{V_{C}-V_{G}}{Z_T^{S-F}} \right) + V_A
\end{equation}
\begin{equation}
\frac{V_{A}-V_{D}}{Z_{C}^S}=\frac{V_{D}-V_{C}}{Z_{DR}^S}+\frac{V_{D}-V_{B}}{Z_L^S}+\frac{V_{D}-V_{H}}{Z_T^{S-F}}
\end{equation}
For simpler calculations, the admittance of each loop is calculated and formulated as the admittance matrix $M_G$ as shown below. 
\begin{equation}
{M}_{G} =
 \begin{bmatrix}
  \sum_{i=1}^{n}\frac{1}{Z_{1i}} & -\frac{1}{Z_{12}} & \cdots & -\frac{1}{Z_{1n}} \\
  -\frac{1}{Z_{21}} & \sum_{i=1}^{n}\frac{1}{Z_{2i}} & \cdots & -\frac{1}{Z_{2n}} \\
  \vdots  & \vdots  & \ddots & \vdots  \\
  -\frac{1}{Z_{n1}} & -\frac{1}{Z_{n2}} & \cdots & \sum_{i=1}^{n}\frac{1}{Z_{ni}}
 \end{bmatrix}
\end{equation}
where $Z_{nm}$ is the impedance between node \textit{n} and node \textit{m}. The current at each point is calculated based on the following relation.
\begin{equation} \label{tension}
M_G.\hat{V} = \hat{I}
\end{equation}
where $\hat{V}$ is the vector with tensions that needs to be found, and $\hat{I}$ is the vector with the sum of currents through each node. From the KCL node equations and the voltage vector $\hat{V}$ and current vector $\hat{I}$ representing the sum of currents entering or leaving node can be represented as 
\begin{equation*}
\begin{aligned}[c]
    \hat{V} =
 \begin{pmatrix}
  V_{1} \\
  V_{2} \\
  \vdots \\
  V_{
  n}
 \end{pmatrix}
\end{aligned}
\qquad\&\qquad
\begin{aligned}[c]
 \hat{I} =
\begin{pmatrix}
 I \\
  0 \\
  \vdots \\
  0
  \end{pmatrix}
\end{aligned}
\end{equation*}
where $V_{n}$ is the voltage at node \textit{n}. The position of I depends on the placement of the source.  The voltage received across any of the branch between C-D, G-H, and so on in Fig.\ref{fig:3d}(d) can be calculated based on the location of the receiver electrodes. The transfer function from the circuit in Fig.\ref{fig:3d}(d) is calculated using 
\begin{equation} \label{ECAgain}
G(w,E_L,D,E_S,[T])=20.log_{10} \bigg| \frac{V_{o}}{V_{I}}\bigg|
\end{equation}
where [T] is the vector of tissue thicknesses for skin, fat, muscle and bone, $V_{o}$ is the potential difference observed across the receiver electrodes and $V_{I}$ is the source voltage. We tracked the phase shift information using the following equation.
\begin{equation} \label{Phase}
\text{Phase}=\text{arctan}\left( \frac{Im(V_o)}{Re(V_o)}\right)
\end{equation}
The channel characteristics computed using the model thus derived are presented and verified in  Section~\ref{Results}. It can be seen from the derivations that the model is more expressive and one can demonstrate the ability to analyze the impact of various network parameters such as electrode size, transmitter receiver separation, and tissue thickness among others on sensor placement and tissue channel performance. 
\section{Simulation Framework for Model Verification}\label{FEM}
In this section, we describe the tissue modeling using the Ansys HFSS, which allows us to perform full-wave electromagnetic simulations for arbitrary 3-D models. It allows detailed computational analysis of field distribution at various locations inside the tissues using finite element analysis (FEA), and is especially useful when experimental results are not easily obtained for intra-body channels. 

We model the forearm with dimensions as described in Section \ref{Sec:Model}. \tc{A pair of copper cuboids of dimension $10 \times 10 \times 1\,\mathrm{mm}$ that is similar to TEC model} is used as the terminal and reference electrodes. The electrodes are connected by a complex impedance defined lumped port. The source current of $1\,\mathrm{mA}$ is set at the lumped port (input). To $1\,\mathrm{foot}$ distance around the forearm model, we emulate a boundary as an open electrical circuit. The frequency dependent electrical properties of dielectric tissue blocks are configured using (\ref{eps_eqn})-(\ref{epsdoubledash}) for the frequency range $100\,\mathrm{kHz}$ to $1\,\mathrm{MHz}$. 

\begin{figure}[t]
 \centering
 \includegraphics[width=7cm,height=4.5cm]{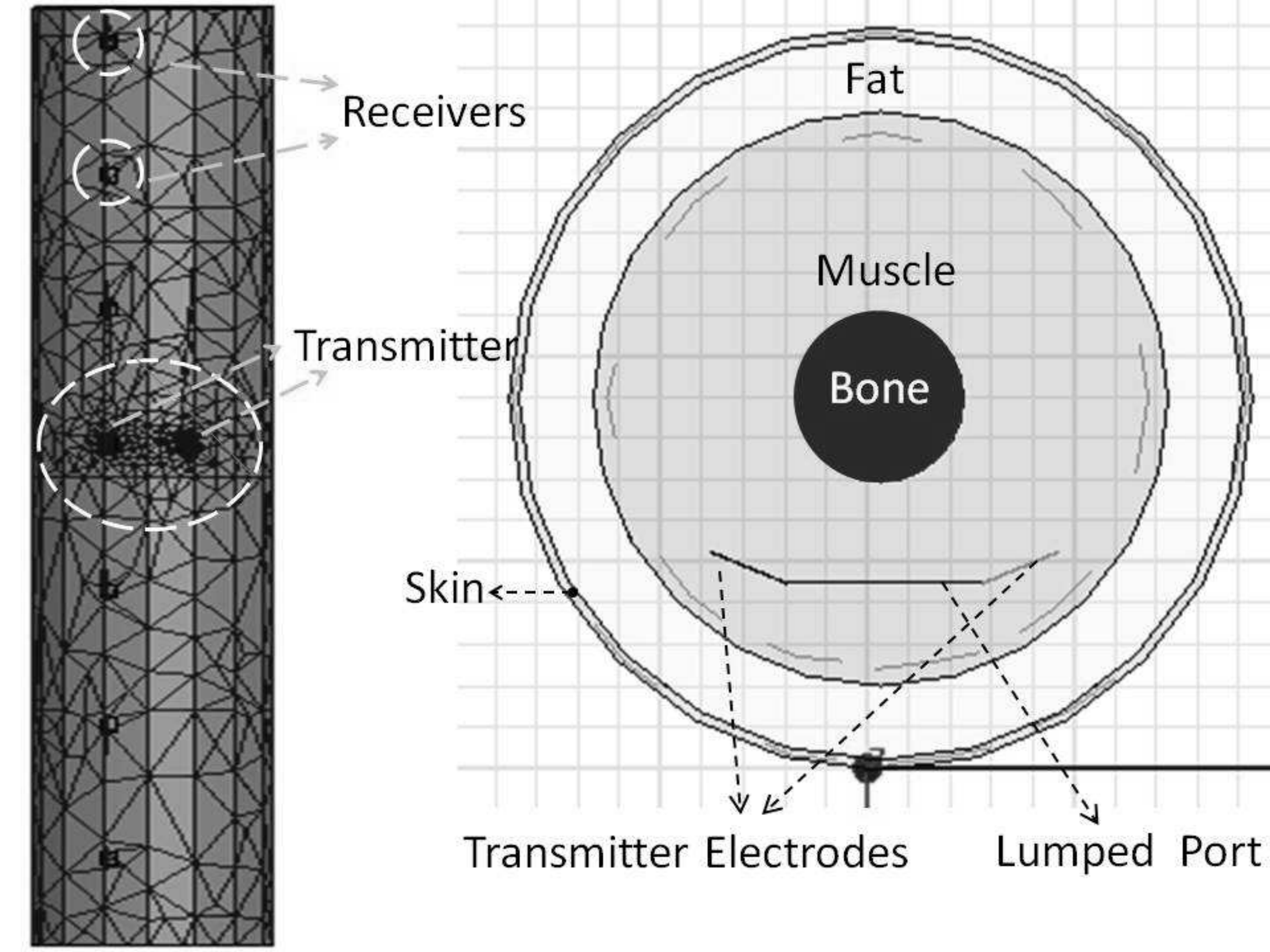}
 \caption{Simulated Forearm Model using FEA; Discretized with high density at critical areas (left); Top view with transmitter electrodes in muscle (right)}
 \label{fig:FEMmodel}
 \end{figure}

HFSS transforms the 3-D tissue model into a mesh of tetrahedron structures, with a high density of mesh points at critical positions like the electrode-tissue interface (Fig.\ref{fig:FEMmodel} (left)). \tc{We performed the analysis in terms of the equivalent electric and magnetic ($E$ and $H$) fields in simulation in contrast to current and voltage ($I$ and $V$) vectors in TEC model to estimate the channel gain.} To determine the field strength across the above said tetrahedrons, complex $EM$ field values at each vertex of tetrahedron is computed using Maxwell's partial differential equations. The normal $E$ component on skin surface is measured as surface integral over an area equivalent to the surface area of a receiving electrode. The $H$ field is measured as surface integral of its tangential component.
The current through surface S at distance \textit{l} from the source can be obtained from Ampere's law as $I_{\perp S} (l) = \oint H.dl$.

\begin{figure}[t]
 \centering
 \includegraphics[width=7.5cm,height=4.5cm]{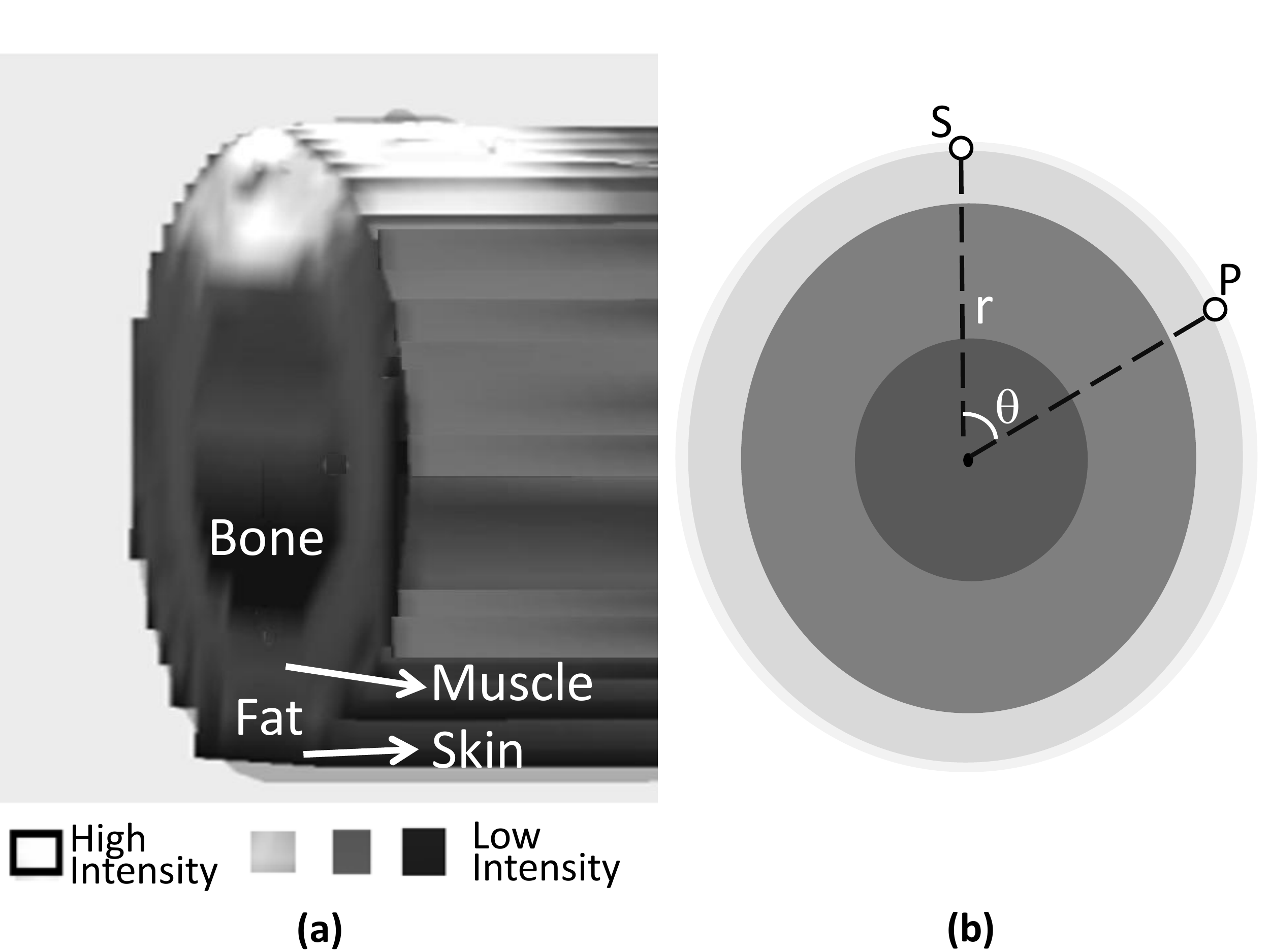}
 \caption{(a) Tissue signal distribution illustrating lower signal strength in bone and higher signal strength in muscle (b) Influence of $r$ and $\theta$.}
 \label{fig:layerprop}
 \vspace{-6mm}
 \end{figure}

\tc{From Fig.\ref{fig:layerprop}.(a), we see that the signal propagates disparately in each layer. For instance, along the lateral direction, the signal propagates only through a part of bone. However, in the muscle, the signal propagates through the entire tissue (refer Fig.\ref{fig:layerprop}.(a)). The signal strength at any point $P$ (refer Fig.\ref{fig:layerprop}.(b)) in a tissue depends on its electrical properties and on the distance between source $S$ and $P$ along the tissue and is independent of the distance from center of the cylinder $(r)$, or the azimuth angle $(\theta)$ between the line connecting center to $P$ and a reference plane. For this reason, we approximate the curvature SP of the cylindrical arm as the Euclidean distance of rectangular tissues in TEC model in Section.\ref{Sec:Model} (Fig.\ref{fig:armcut}.(b)). In order to achieve model conformance in the FEA cylindrical arm model, we estimate the angle of electrode separation, $\theta$ as $E_S/r$, where $E_S$ is the Euclidean distance of electrode separation in TEC model.} For emulating the signal received at the implanted sensor, we move the transmitter electrodes and port into muscle tissue (Fig.\ref{fig:FEMmodel}(right)). The $E$ field strength measured across the receiver electrodes is used to calculate the output voltage.
The gain through the tissues can be calculated as follows. 
\begin{equation}
\label{E gain}
G_E (dB)= 20\ log_{10} \left(\frac{E_{Detector}}{E_{Coupler}}\right)
\end{equation}
%and
%\begin{equation}
%\label{H gain}
%G_H (dB)= 20.\ log_{10} \left(\frac{H_{Detector}}{H_{Coupler}}\right)
%\end{equation}
The simulation is repeated for different $E_S$ (distance between the terminal and reference electrodes), and D (different distances between the transmitter and receiver) for varying [T] (thickness of tissues) at frequencies ranging from $100 \mathrm{kHz}$ to $1 \mathrm{MHz}$. The results are used to verify our TEC model as discussed in Section~\ref{Results}. \tc{In addition, using the FEM model, we derive the boundary conditions next that are necessary to ensure tissue safety.}
\begin{figure}[b]
 \centering
 \includegraphics[width=9cm, height=5.5cm]{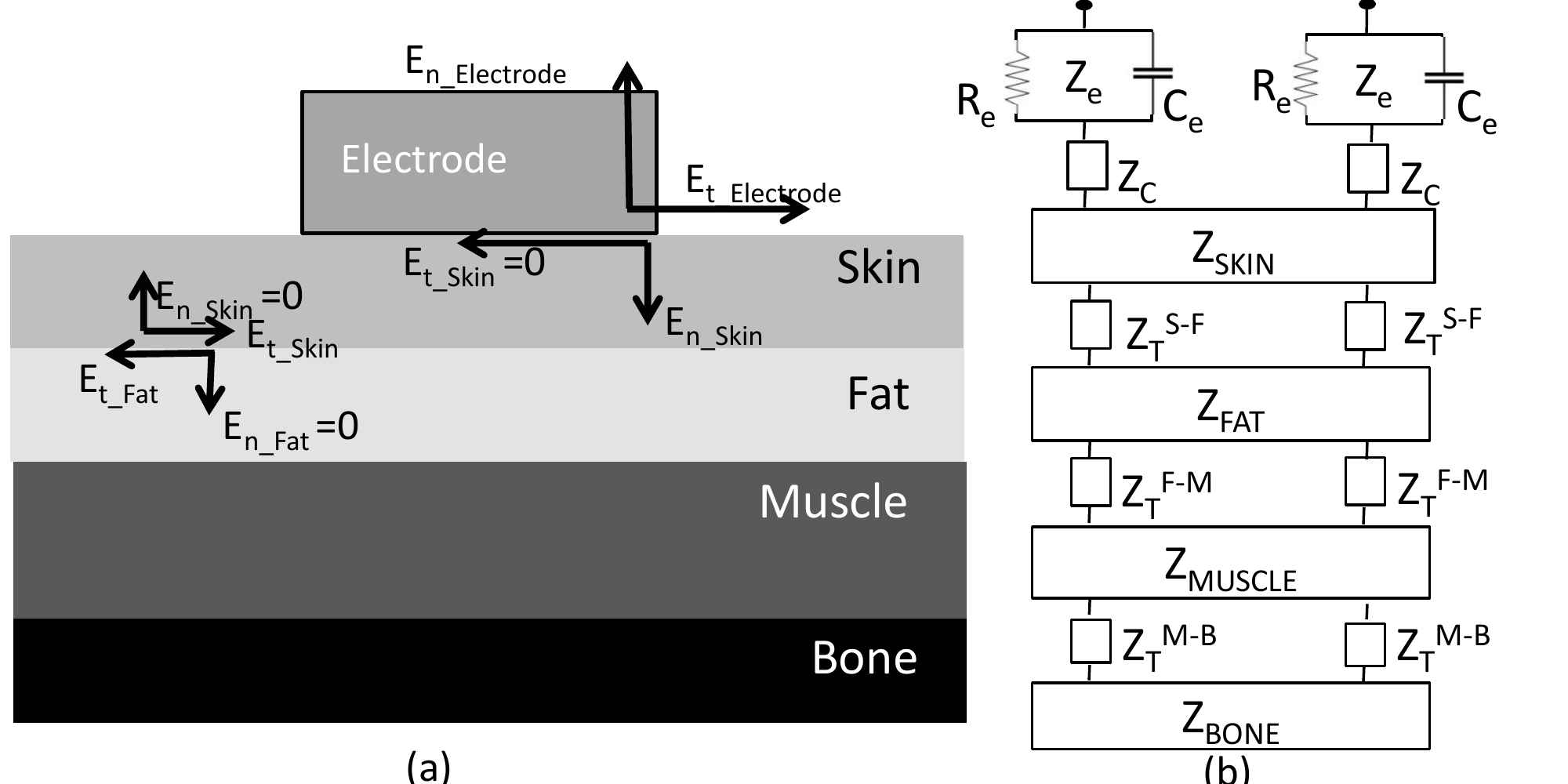}
 \caption{(a) E components at electrode-skin and skin-fat interface (b) Circuit model for interfaces at transmitter side}
 \label{fig:electrode}
 \end{figure}
\subsection*{Ensuring Safe Signaling for Human Tissues \label{sec:safety}}
The energy absorption in tissue is proportional to the conductivity of the medium. At lower frequencies such as $100\,\mathrm{kHz}$ and $1\,\mathrm{MHz}$, conductivity and therefore, the absorption is low resulting in less than $1$ degree temperature rise and no impact on live tissue. Further, the International Commission on Non-Ionizing Radiation Protection (ICNIRP) guidelines \cite{safety,ICNIRP} limits the current density through the human body to $25 \mathrm{mA/m^2}$ in the frequency range of $100\,\mathrm{kHz}$ through $60\,\mathrm{MHz}$ for the general public exposure. In order to ensure that the induced current density in the model is safe for the human tissue, we perform the following analysis. 

On exciting the electrodes with voltage $V$, the potential difference inside the high conductive electrode $\oint E.dl$ becomes zero. The total current flowing through the surface of electrode of uniform cross section is given by
\begin{equation}
I=\int\!\!\!\int_s J.ds,
\end{equation}
where $J$ is the current density. The electric field $E$ at the electrode surface can be decomposed into normal and tangential components as $E_{t\_Electrode}$ and $E_{n\_Electrode}$, where $E_{Electrode}$ = $E_{t\_Electrode} + E_{n\_Electrode}$. At the equipotential electrode-tissue (conductor-dielectric) contact area, the tangential component of electric field $E_{t\_Electrode}$ approaches zero \cite{bioEM-book,biofields} and the non-zero normal component becomes the field of excitation in the tissue given as:
\begin{equation}
E_{n\_Electrode} = E_{n\_Tissue} 
\end{equation}
The $E_{n\_Tissue}$ is shown as $E_{n\_Skin}$ in Fig.\ref{fig:electrode}. At any instance, the current density at the contact area will be the largest among all other parts of tissue as the signal flows radially away from the region of source and attenuates with distance. Therefore, the source region is the area, where the safety levels of current injection is to be confirmed to avoid tissue damage. To ensure the safe limit of exposure in the contact area of dimension $10\,\mathrm{mm} \times 10 \,\mathrm{mm} \times 1 \,\mathrm{mm}$, we limit the current flowing through the tissue at the electrode contact area in such a way that, 
\begin{equation}
I_{contact-area} = \int\!\!\!\int_s J.ds = \int\!\!\!\int_s \sigma . E.ds, \ \leq 1 mA
\end{equation}
where s is the surface area of electrode and $J$ is the current density is given by,
\begin{equation} \label{eqn:J}
J=(\sigma + j\omega \epsilon'')E
\end{equation} 
that includes both conduction and displacement currents.

We confirmed the safe current density level using simulation by measuring the magnitude of current density at the rectangular region in contact with  source electrode (region of maximum exposure). For an input current of $1 \mathrm{mA}$ at 0.5 V, the observed value of $J$ is $0.6\, \mathrm{mA/m^2}$ which is well below the safe limit. In case of multiple transmitters in IBN, the transmitters should be spatio-temporally separated in order to ensure that the cumulatively aggregated values of current density (due to multiple sources) does not exceed beyond the safe level. 
\section{Model Verification \& Discussion}\label{Results}
This section verifies the analytical model derived in (Section~\ref{Sec:Model}) using the simulator design from Section~\ref{FEM}, as well as with prior experimental measurements for S-S path in literature. We use the clinical trial findings described in the existing work \cite{Gal8} and measurements in \cite{distri} for verifying the channel gain obtained through the S-S path and \cite{QS1} for verifying the effect of varying the transmitter-receiver separation distance (D) on gain in M-S path. We conduct the evaluations on the following basis at different paths: (i) variation of gain with frequency, (ii) phase shift of the signal with frequency, and (iii) impact of frequency on energy dissipation. 

\begin{figure}[b]
 \centering
  \includegraphics[width=8.5cm,height=5.5cm]{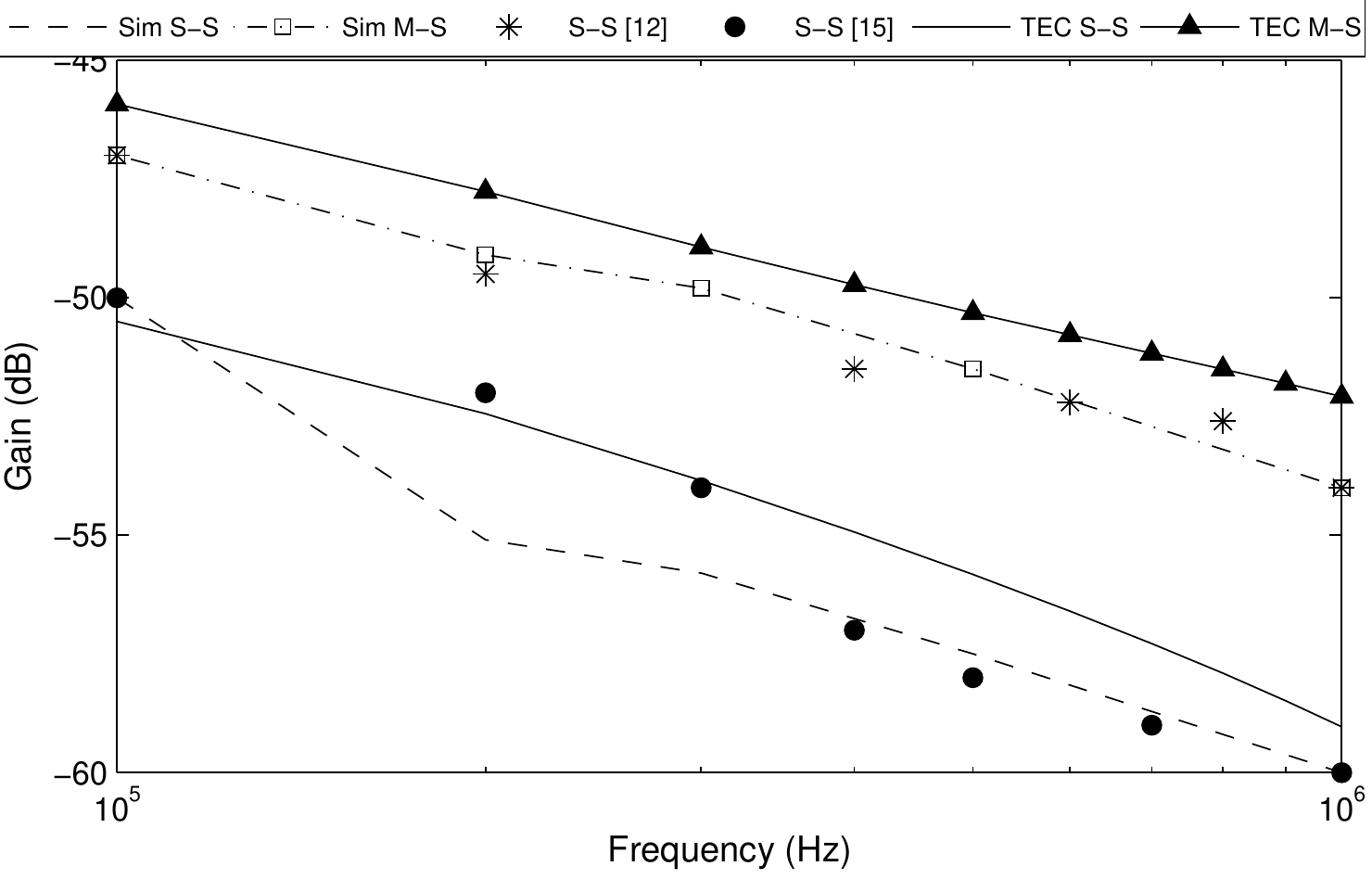}
 \caption{S-S and M-S gain Vs frequency using tissue equivalent circuit model (TEC), simulation (FEA) and literature measurements; D=10 cm, $E_{ST}$=$E_{SR}$=5 cm}
 \label{fig:ss}
\end{figure}

\begin{figure}[b]
 \centering
  \includegraphics[width=9cm,height=5.5cm]{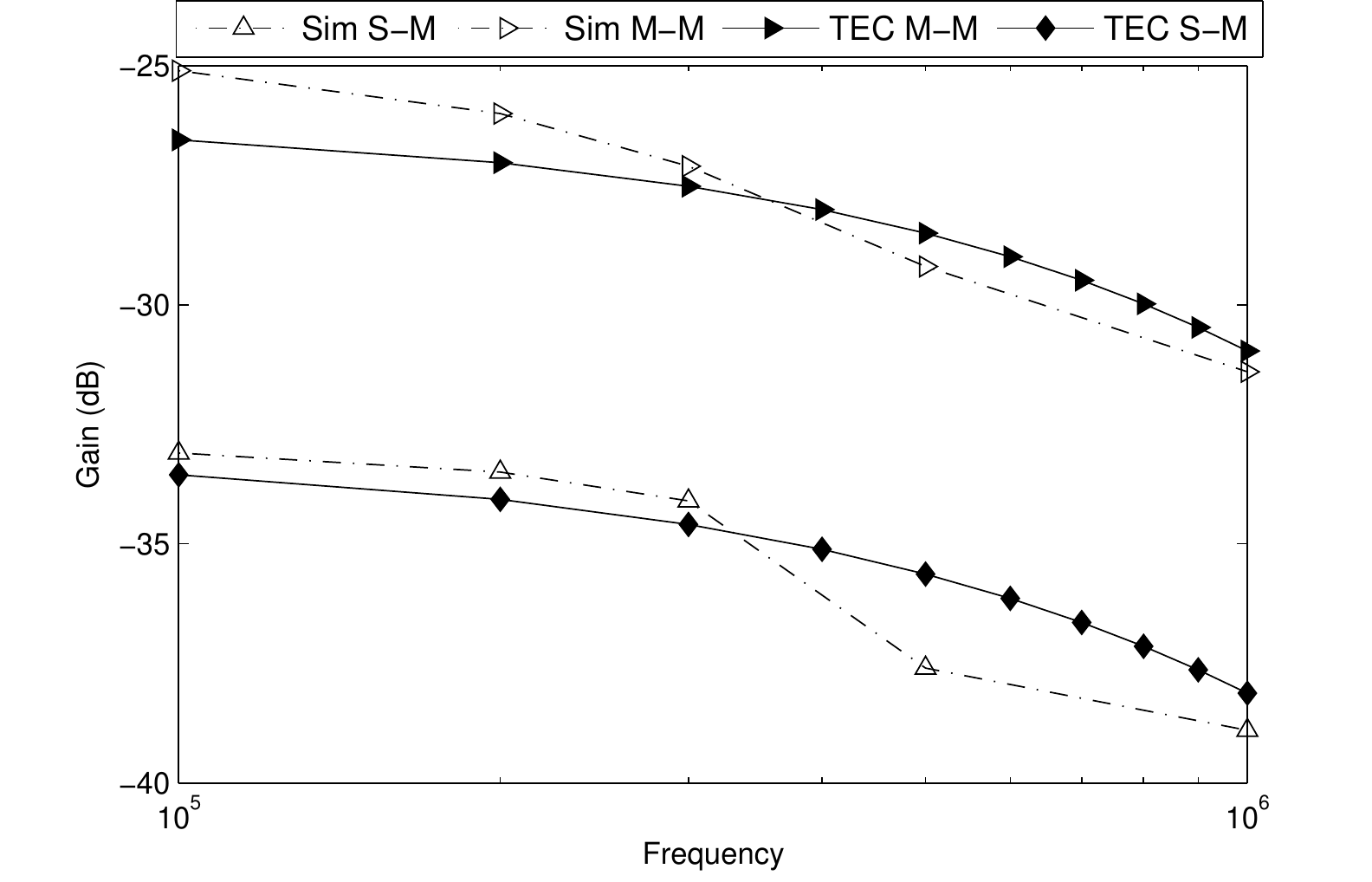}
 \caption{S-M and M-M gain Vs frequency using tissue equivalent circuit model (TEC), simulation (FEA) and literature measurements; D=10 cm, $E_{ST}$=$E_{SR}$=5 cm}
 \label{fig:mm}
\end{figure}

The channel gain obtained from 100 kHz to 1 MHz with D being $100\,\mathrm{mm}$ and the electrode separations in transmitter $E_{ST}$ and receiver $E_{RT}$ being $50\,\mathrm{mm}$ using TEC model (\ref{ECAgain}) and simulation model (\ref{E gain}) are presented for the S-S and M-S in Fig.$\ref{fig:ss}$, and for the S-M and M-M paths in Fig.\ref{fig:mm}. The tissue dimensions are specified in section.\ref{Sec:Model}. The values we choose for $F_W$, m and m' are $0.7, -1.15$ and $-0.81$ \cite{bodyimpedance}. The channel gain obtained using TEC model (Fig.\ref{fig:ss}) at $100\,\mathrm{kHz}$ is around $-50\,\mathrm{dB}$ and drops by $10\,\mathrm{dB}$ at $1\,\mathrm{MHz}$ on the S-S path. We see good agreement among the TEC and simulation model plots and with prior experimental results from literature for the S-S path. The variation between the TEC model results and simulation results is less than $2\,\mathrm{dB}$, verifying the accuracy of the model. The channel gain obtained for TEC model S-S path matches well with the clinical trails in \cite{Gal8}, where the electrodes and tissue dimensions used are similar to the ones assumed in our analysis. 

\tc{There is a difference of about $3\,\mathrm{dB}$ with the measurements from \cite{distri}, which we attribute to the variation in the electrode dimension (circular electrode with radius $0.5\,\mathrm{cm}$) and the usage of electrode conductive gel. There are other inherent measurement uncertainties associated with GC-IBN including tissue temperature, hydration levels and surface treatment that we capture using parameters $F_W$ and $Z_{Co}$ for a typical adult, which are not specified in \cite{distri}. Moreover, the literature reports a variation of $2\,\mathrm{dB}$ among measurements on different days. The above mentioned reasons along with variation in $\sigma$ and $\epsilon$ values of tissues among individuals by $\pm 0.1\,\mathrm{S/m}$ and $\pm 0.05$ respectively, in the range of frequency used \cite{prop1}  contribute to the difference between our results and those reported in \cite{distri}}.

We observe that the gain obtained in the muscle tissue is significantly higher than the S-S path by $\approx 24\,\mathrm{dB}$ advantage in gain with $-26\,\mathrm{dB}$ at $100\,\mathrm{kHz}$, that drops by $\approx 4\,\mathrm{dB}$ at $1\,\mathrm{MHz}$, indicating better SNR and less frequency sensitivity in M-M path. Note that the S-S path gives a gain variation of $\approx 10\,\mathrm{dB}$ in the range of frequency considered. The S-M and M-S paths have channel gain higher than the S-S path but lower than the M-M path. The S-M path with the receiver placed in muscle has atleast 12 dB more gain than the M-S path with the receiver on skin. As there are no published experimental data on the signal gain over the M-M, S-M $\&$ M-S paths to our best knowledge, our studies are limited to comparison between the analytical and theoretical models we have derived in this work. 

\noindent $\bullet$ \textbf{Phase shift of the signal with frequency:}
We next study the impact of tissue channel on the transmitted signal phase using (\ref{Phase}), at S-S, S-M, M-S and M-M paths. Fig.\ref{fig:5plot}(a) shows the shift in phase when the signal frequency varies in the range of $100\, \mathrm{kHz} - 1\, \mathrm{MHz}$. 
\begin{figure}[t]
 \centering
\includegraphics[width=8cm,height=4cm]{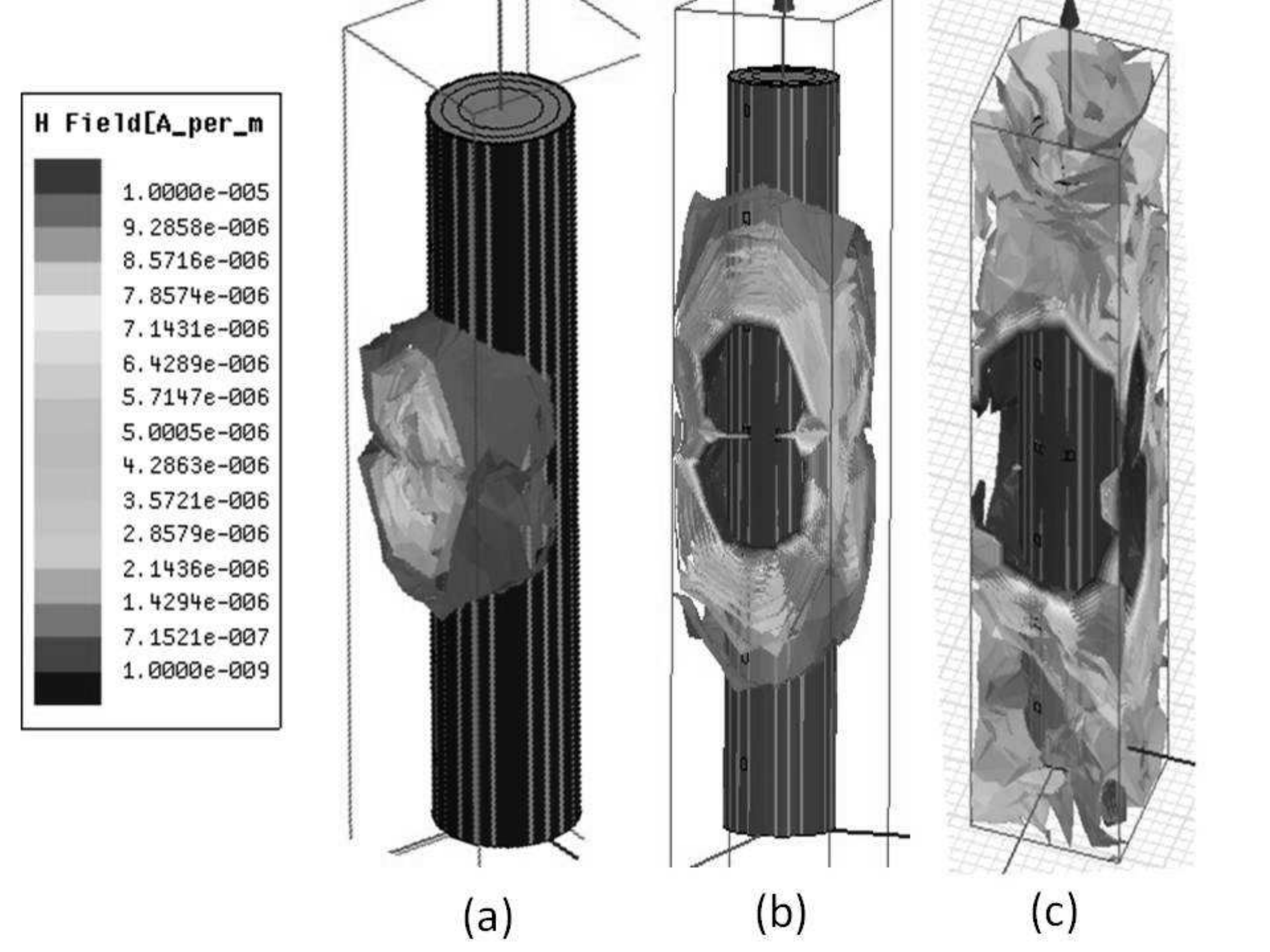}
 \caption{H Field spreading out of body at (a) $100\,\mathrm{kHz}$ (b) $1\,\mathrm{MHz}$ (c) $10\,\mathrm{MHz}$}
 \label{fig:field}
 \end{figure}
We observe that the phase shift on the S-S and M-S path varies from 16 to 20 degrees, whereas for  the M-M and S-M paths, there is less than 7 degrees of shift in phase reinforcing that the muscle tissue serves as a better channel.

\noindent $\bullet$ \textbf{Impact of operating frequency:} 
To identify the ideal range of the transmission frequency, we consider two factors: (i) frequency of the signals naturally generated by the human body, and (ii) signal loss caused by dissipation for a given frequency within the tissue. The electrical signals within the human body including neural impulses, ECG, and EEG signals operate at a frequency lower than $50\,\mathrm{kHz}$, and therefore, we avoid the frequencies $\le 50\,\mathrm{kHz}$ for intra-body communication. As the channel characteristics are frequency dependent, we need to identify the ideal operating frequency that reduces signal loss.

The signals transmitted into the tissue results into two current components, i.e., the conduction current and displacement current as given in (\ref{eqn:J}). At lower frequencies, the conduction current that is caused by the movement of charges is high. This enables energy detention inside the tissue, resulting in higher intensity at the receiver end. At higher frequencies above $1\,\mathrm{MHz}$, the conductivity remains constant and therefore the conduction current also remains fixed. However, due to increase in capacitance effect the displacement current grows larger with frequency. This ultimately results in signal dissipating from the body into the surrounding region, possibly causing interference externally, as well as limiting the energy incident on the receiver electrode. 

For instance, at $100\,\mathrm{kHz}$, the H field in the surrounding the body is in the order of a few ${\mu A/m}$, extending to around $50\,\mathrm{mm}$ at the exterior. On the other hand, at $10\,\mathrm{MHz}$, the H field surrounding the body is higher by two orders of magnitude, extending to about $3$ feet away from the body (refer Fig.\ref{fig:field}). The signal spreading out of the body is considered wasted, as it cannot reliably be detected at the embedded receiver. Thus, the signal loss is minimized as long as the operating frequency is restricted in such a way that the conduction current dominates the displacement current. This is true when the relationship $\displaystyle \frac{\sigma}{w\epsilon''} > 1$ holds in all tissues, i.e., when we limit the frequency lower than $2\,\mathrm{MHz}$. Thus, to ensure that the dissipation loss is at minimum, the maximum frequency of operation is set at $1\,\mathrm{MHz}$.
\section{Model Sensitivity Analysis}\label{Sec:sense}
The model proposed in this paper uses different variables as network design parameters such as tissue thicknesses, transmitter-receiver separation, electrode dimensions, and terminal separations. A better understanding of the relationships between these parameters and the channel gain would help determining the placements of IBN nodes. For this purpose, we under take one-factor-at-a-time approach to study the influence of the key network parameters on channel gain in this section. 
\subsection{Effect of Tissue Thickness on Channel Gain}\label{thick}
One of the important parameters that determine channel gain is the thickness of each tissue layer. In this section, we investigate the impact of fat and muscle tissue thickness on the signal gain.  As sensors are often placed either on the skin (with non-invasive access) or in the muscle (best propagation characteristics), the intermediate fat tissue behavior and its thickness play a crucial role in determining the quantity of signal that transcends the tissue boundaries. For instance, the influence of tissue thickness as a parameter in transverse impedance $Z_T$ of the model is given by 
\begin{figure*}[t]
 \centering
\includegraphics[width=18cm,height=8cm]{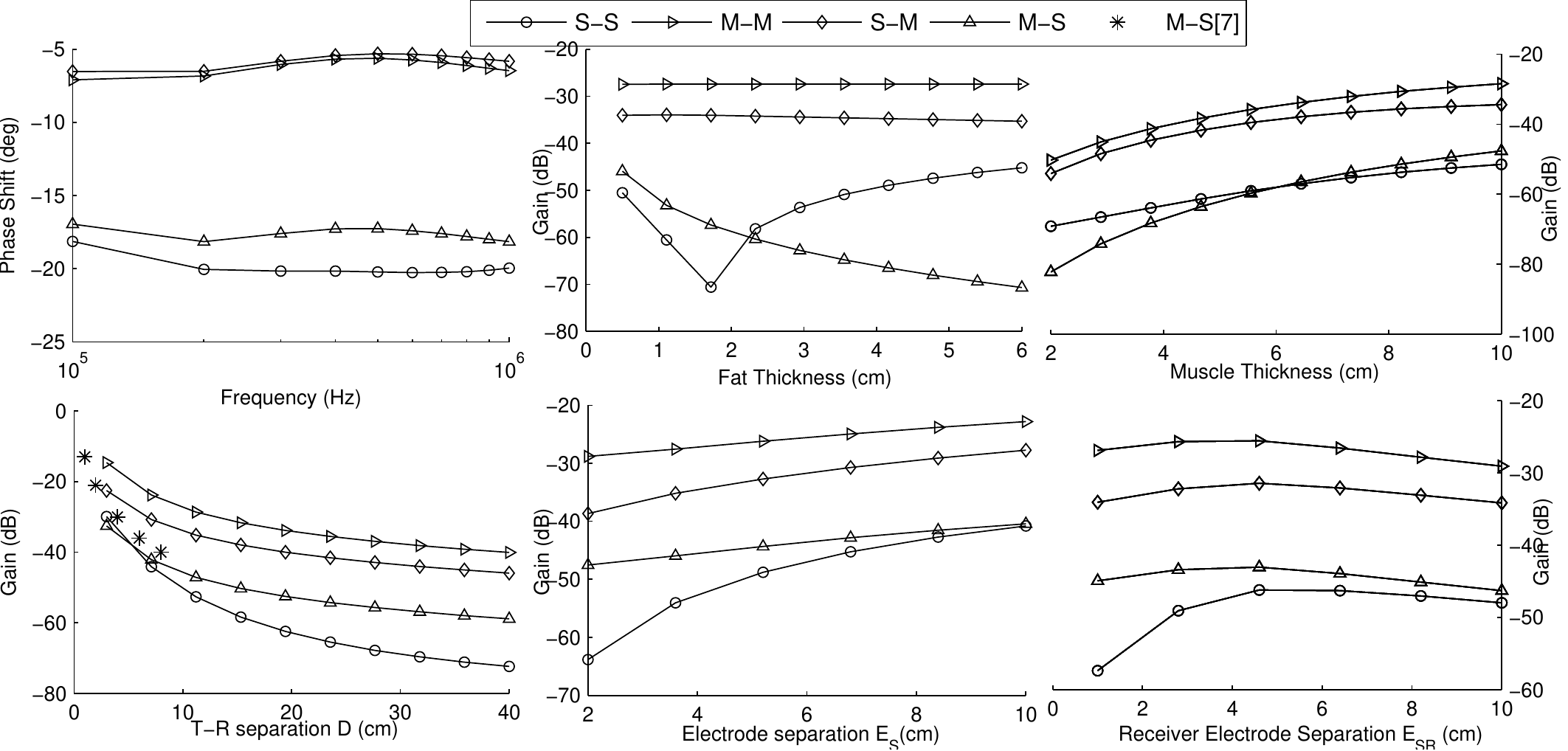}
 \caption{Sensitivity Analysis. Top row: (a)Phase Shift (left) Vs Frequency (b) Gain Vs Fat thickness (c) Gain Vs muscle thickness. 
Bottom row:(d) Gain Vs transmitter - receiver separation (D) (e) Gain Vs electrode separation in transmitter \& receiver ($E_S$) (f) Gain Vs electrode misalignment ($\Delta \ell$)}
 \label{fig:5plot}
 \end{figure*} 
\begin{equation}
Z_T=\frac{(T+\gamma)(\rho + i \omega\epsilon)}{\rho E_L^2 ( \rho + 2i \omega \epsilon)} 
\end{equation}
where $\gamma $ denotes the change in tissue thickness from the average value considered in this paper. In general, fat acts as a barrier between skin and muscle tissues, allowing either tissue to retain the energy (for $\gamma > 0$) or allowing more current to pass through (for $\gamma < 0$). For the channel gain results given in Fig.\ref{fig:ss} and Fig.\ref{fig:mm}, we considered an average value of forearm fat thickness as 7 mm. From the results of varying fat thickness in Fig.\ref{fig:5plot}(b), it can be seen that for varying fat thickness from $0.5\,\mathrm{mm}$ to $60\,\mathrm{mm}$ with $D=100\,\mathrm{mm}$, $E_S=50\,\mathrm{mm}$ at $100\,\mathrm{kHz}$ the M-M path shows no significant change in gain and performs better for all fat thicknesses assumed. The S-M path has a slight drop in gain by about 1 dB illustrating that for any fat thickness, the dominant part of signal propagates through the muscle. The M-S path gain also drops with fat thickness when there is no signal leakage from muscle to skin for thick fat. The S-S path gain drops for fat thickness between $1$ and $3\,\mathrm{cm}$ and then improves towards the thin fat values when there is minimal leakage to the layers beneath the skin. 

We can conclude that for a thick fat layer, the receiver should be positioned in the same tissue layer as the transmitter for better channel gain. As signal leakage is non-negligible for any fat thickness, simultaneous communication on the skin and within the muscle cannot coexist at the same frequency. Thus for multiple pair of co-located sensors and actuators placed on the skin as well as implanted within the muscle to be active, a multi-access scheme is required. For covering longer distances, and if the BMI values indicate thick fat layer, the M-M path is preferable. We undertake a similar study for varying muscle thickness and the results are given in Fig.\ref{fig:5plot}(c). The gains along all the four considered paths increases with muscle thickness. In M-M, M-S and S-M paths, for $40\, \mathrm{mm}$ increase in muscle thickness, the increase in gain is about $15\, \mathrm{dB}$ while in S-S path, the gain increases by $8\, \mathrm{dB}$. Thus, networks formed in thicker muscle tissue offer better channel gains and cover longer distances. 
\subsection{Impact of Transmitter-Receiver Separation Distance}
The maximum possible transmitter-receiver separation distance $(D)$ that determines the quality of signal for communication is one of the primary factors in IBN design. Transmitted signals suffer a natural attenuation with distance owing to the increasing longitudinal impedance, $Z_L$. Using analytical model, the impact of variation in $D$ in the longitudinal and cross impedance (Fig.\ref{fig:sense}(a) can be derived in terms of the network parameters considered in this section as,
\begin{equation}
Z_L=\frac{D (T \rho + i\omega \epsilon D )}{A\rho(T \rho + 2i\omega \epsilon D)} 
\end{equation} and
\begin{equation}\label{eqn:zc}
z_C=\frac{\sqrt{2(D^2+E_S^2)}(T^2 \rho + i \omega \epsilon(D^2+E_S^2))}{2 \rho E_L T(T^2 \rho + 2i \omega \epsilon (D^2+E_S^2))} 
\end{equation}
The rate of change of $Z_L$ with respect to the change in $D$ is inversely proportional to $D$ that reflects similar trend in the channel gain calculation as illustrated in Fig.\ref{fig:5plot}(d). For an increase in $D$ from $20$ to $100\,\mathrm{mm}$, the signal gain drops by around $18\,\mathrm{dB}$ in S-S path, about $10\,\mathrm{dB}$ in M-M path, and about $12\,\mathrm{dB}$ in S-M/M-S paths. This analysis would help determine the single-hop distance in body network design. 
\subsection{Impact of Electrodes Separation Distance}
\begin{figure}[t]
 \centering
 \vspace{-2mm}
\includegraphics[width=8.5cm,height=5.2cm]{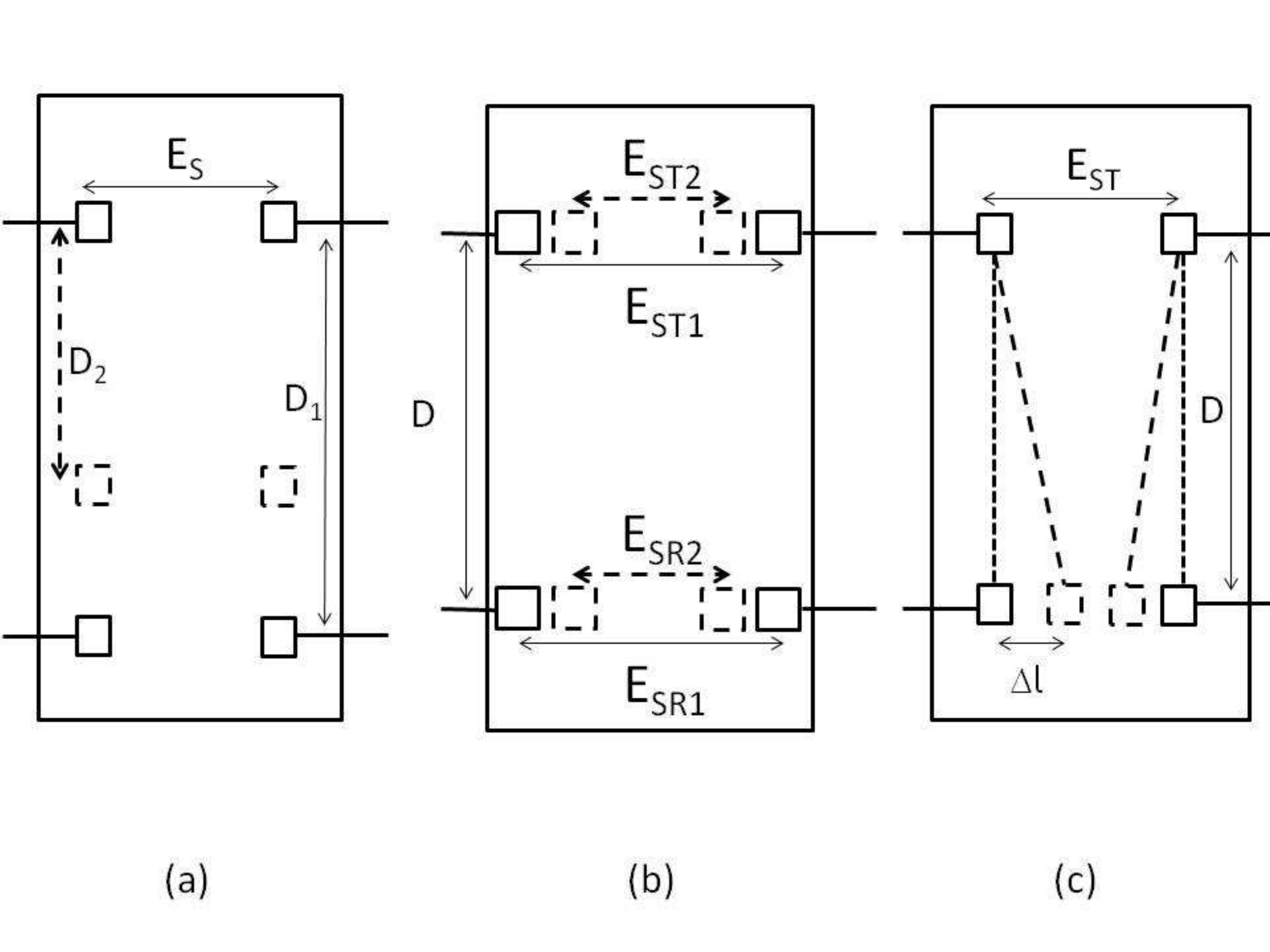}
 \caption{Electrode placements for various (a)Transmitter - receiver separation (b) Electrode separation (c) Transmitter - receiver electrodes alignment }
 \vspace{-4mm}
 \label{fig:sense}
 \end{figure}

Fig.~\ref{fig:sense}(b) illustrates variation in the electrode separation distance, $E_S$ of the transmitter and the receiver together. The effect of $E_S$ is prominent on the direct impedance $Z_D$ as given by the following relation.
\begin{equation} \label{eqn:zd}
z_D=\frac{ E_S(D^2 \rho +i \omega \epsilon E_S^2)}{\rho E_L  D(D^2 \rho + 2i \omega \epsilon E_S^2)} 
\end{equation}
The gain in all paths increases with $E_S$ as shown in Fig.\ref{fig:5plot}(e). Moving the electrodes far apart, such as for the separation achieved by positioning one electrode on the top surface and the other one on the bottom surface of the forearm, the gain dramatically increases to a maximum of $25\,\mathrm{dB}$. We observe similar trends when the separation distance is varied within muscle (i.e., the M-M case). For instance, by parting the electrodes from $20\,\mathrm{mm}$ to $100\,\mathrm{mm}$, the increase in gain is about $20\,\mathrm{dB}$ in S-S path, $5\,\mathrm{dB}$ in M-M path, and $8\,\mathrm{dB}$ in S-M and M-S paths for average fat width.
\tc{\begin{table}[b]
\centering
\caption{\label{tab:porksimilarity} Similarity in electrical properties of Porcine (P) \& human(H) tissue}
\begin{tabular}{c|c|c}
\hline
Tissue	& Conductivity (S/m) &	Permittivity\\ 
\hline
Skin (dry)	&  0.00016 (P)	& 965 (P)\\
&	0.00045 (H) &	1119.2 (H)\\
\hline
Fat	& 0.03 (P)	& 98 (P)\\
&	0.024 (H)	& 92.8 (H)\\
\hline
Muscle	& 0.25 (P) 	& 9900 (P)\\
&	0.36 (H)	& 8089.2 (H)\\
\hline
\end{tabular}
\end{table}}
\subsection{Effect of Transmitter and Receiver Alignment}
In the above discussion, we considered equal distances between the electrodes of transmitter $E_{ST}$ and  receiver $E_{SR}$ with the transmitter electrode pair perfectly aligned with that of receiver along the longitudinal direction as shown as dotted line in Fig.\ref{fig:sense}(c). In this section, we assume the possibility of electrodes' mis-alignment shown as dashed lines in Fig.\ref{fig:sense}(c) deviated by $\Delta \ell$ from aligned position and study its impact on the channel gain. $\Delta\ell$ shown in Fig.\ref{fig:sense}(c) illustrates only the position in-between the dotted lines that would reduce $E_{SR}$ while it can also be a deviation outside the dotted lines that would increase $E_{SR}$ further. The following equation shows the modified expression for $Z_L$ that includes the influence of  mis-alignment $\Delta\ell$.
\begin{equation} \label{eqn:misalign}
Z_L=\frac{\sqrt{(D^2+\Delta\ell^2)} (T \rho + i\omega \epsilon \sqrt{(D^2+\Delta\ell^2)}}{A\rho(T \rho + 2i\omega \epsilon  \sqrt{(D^2+\Delta\ell^2)})} 
\end{equation}
It is found that the gain decreases with $\Delta \ell$ caused by the increase in $Z_L$ as shown in (\ref{eqn:misalign}) and in other impedance irrespective of the direction of deviation (inside or outside). Maximum gain is obtained for the perfect alignment ($\Delta \ell = 0$) as observed in Fig.\ref{fig:5plot}(f).
\subsection{Electrode dimensions:} The electrode size specified by $E_L$ also has same effect as that of electrode separation, $E_S$. It can be seen from the impedance relationships given in (\ref{eqn:zc}) and (\ref{eqn:zd}) that larger electrode dimensions could lead to higher gain. For instance, an increase of $10\,\mathrm{mm}$ in $E_L$ of electrode brings in $8\,\mathrm{dB}$ of improvement in gain. However, larger on-skin or implanted nods may cause discomfort. Thus, a compromise between electrode size and gain can help decide the transmitter - receiver distance, the need for next hop relay nodes and their best possible location. 
\section{Model validation using Porcine Experiments}\label{Sec:pig}
In addition to the verification of the proposed TEC model using simulation and literature measurements, we also performed empirical validation of our model using galvanic coupled channel gain measurements with porcine tissue as the transmission medium. \tc{The porcine tissue is considered for validating our analytical model because of the similarities between human and porcine tissues with respect to cutaneous blood supply, body surface areas, cellular turnover rate ($28 - 30$ days), lipid composition and also in their electrical properties. The porcine electrical properties match accurately with the Cole-Cole model \cite{pigprop}. Table.\ref{tab:porksimilarity} illustrates the similarity in electrical properties between human and porcine tissues.} The analytical model was adapted to the electrical properties of the porcine tissue used in \cite{pigprop,prop1,pprop,pprop1}. 
%skin thickness - 2.5 mm for human and 2.2 mm for porcine 
%skin impedance:
%at 100 kHz
%cross impedance - 180 ohms
%longitudinal imp - 80 ohms
%direct imp - 190 ohms
%At 300 MHz, 
%$epsilon_r=50$
%sigma=0.45
\subsection{Measurement Set-up and Calculation}
\vspace{2mm}
The porcine tissue sample obtained from a local slaughter house was extracted with skin, fat and muscle on from a pig weighing 260 pounds. Samples of dimension $34\times 25\times 5\ \mathrm{cm}^3$ were cut from the loin surrounding the hip bone and immediately used for our experiments. %The measurements were taken at the site of sacrifice within an hour of excision to avoid drastic changes in electrical properties from original in-vivo values. 
\tc{To ensure fixed and tight holding on the irregular tissue surface, we used the alligator clips (40 $\mathrm{mm}$) as the electrodes at the two transmitter terminals and two receiver terminals. We modified the electrode material and dimension accordingly and removed the bone layer in TEC model to enable results comparison.}

\begin{figure}[t]
 \centering
\includegraphics[width=8cm,height=5cm]{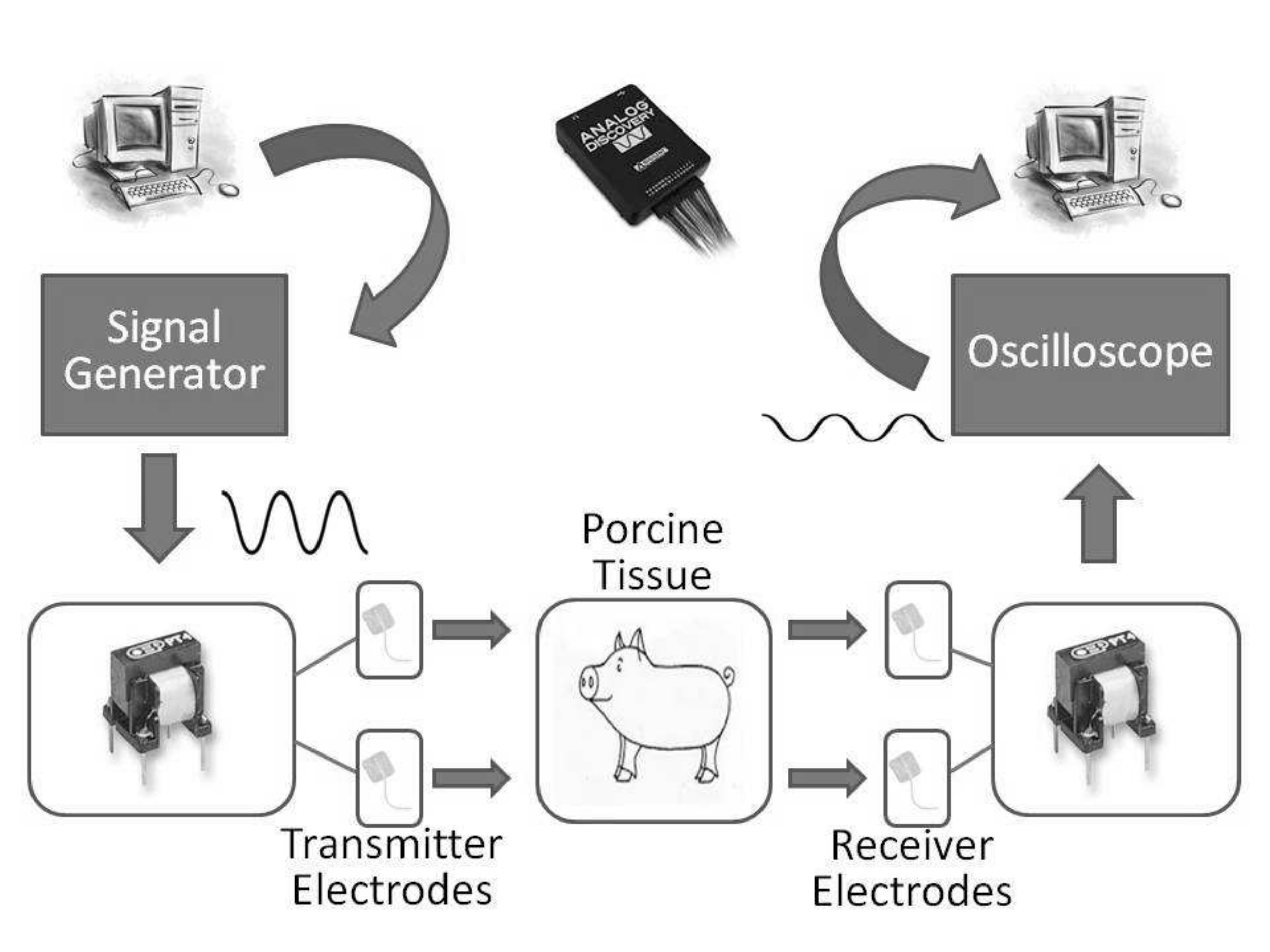}
 \caption{Block diagram for galvanic coupling with porcine tissue}
 \label{fig:pigblock}
 \end{figure}

\begin{figure}[t]
 \centering
\includegraphics[width=8cm,height=5cm]{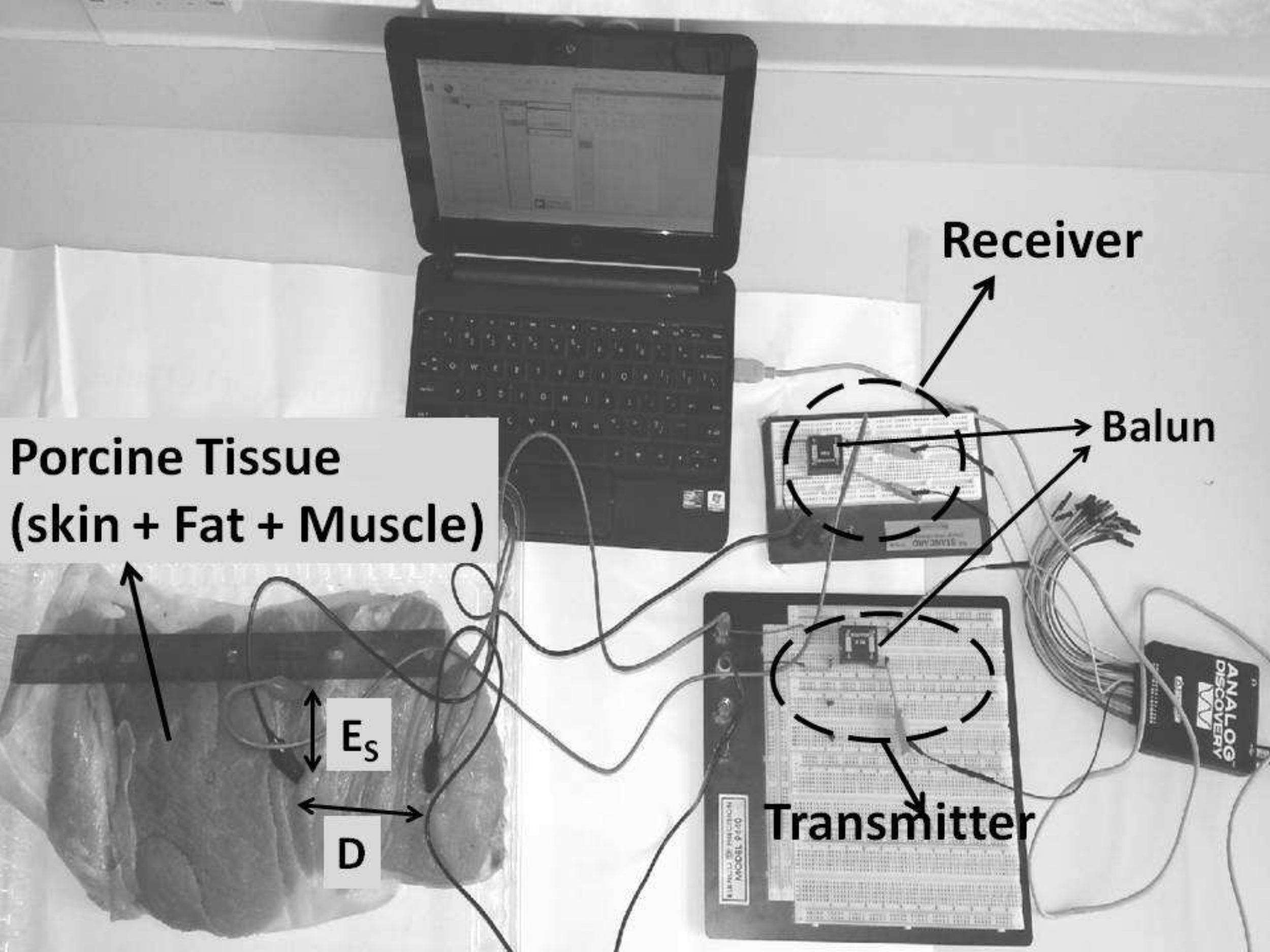}
 \caption{Experimental set-up for galvanic coupling with porcine tissue}
 \label{fig:pigexpr}
 \end{figure}

The skin was cleaned, slightly abraded and moistened on the location where the electrodes are to be attached. A portable bi-channel signal generator and oscilloscope were used for carrying out the experiment on-site. The block diagram for the basic connection and the actual experimental set-up are shown in Fig. \ref{fig:pigblock} and Fig.\ref{fig:pigexpr} respectively. For isolating the transmitter and receiver, we used the OEP PT4 1:1 pulse transformers, one in between the signal generator and transmitter electrodes, and the other in between the receiver electrodes and oscilloscope. 

\tc{Initially, we connect the input directly to the receiver and measure the signal across the receiver terminals without the tissue channel in-between to read the attenuation through transmitter and receiver electronics and the noise. We then introduce the tissue channel, and note the loss incurred through the tissue at $100\,\mathrm{kHz}\text{ and at } 1\  \mathrm{MHz}$. We extract the path loss through tissues using the transmitted signal strength, the received signal strength and the obtained channel attenuation.}
%\textcolor{red}
%{We use the signal averaging technique to mitigate the white and electronic noise encountered in the experiment. }

We approximate the observed instantaneous ambient and electronic noise as Gaussian distribution with zero mean. The effect of noise is mitigated using time average of
% not ensemble assuming it is statistically stationary 
the periodic signal over $10^6$ signal oscillations as 
\begin{equation}
\overline{V_o^2(t)}=\frac{1}{T}\int_{t-T/2}^{t+T/2} V_o^2(t)dt 
\end{equation}
where ` $\bar{\  }$ ' notation denotes the time averaged signal and $T$ is the signal duration %varied between $10\,\mathrm{sec}$ and $1\,\mathrm{sec}$ 
for operating frequency between $100\,\mathrm{kHz}$ and $1\,\mathrm{MHz}$.
%The environmental noise observed with tissue channel cannot be completely mitigated through averaging. Suitable mechanisms to estimate the instantaneous channel characteristics that could nullify the noise and account for the tissue properties uncertainty are required.
\begin{figure}[b]
 \centering
\includegraphics[width=8cm,height=4.5cm]{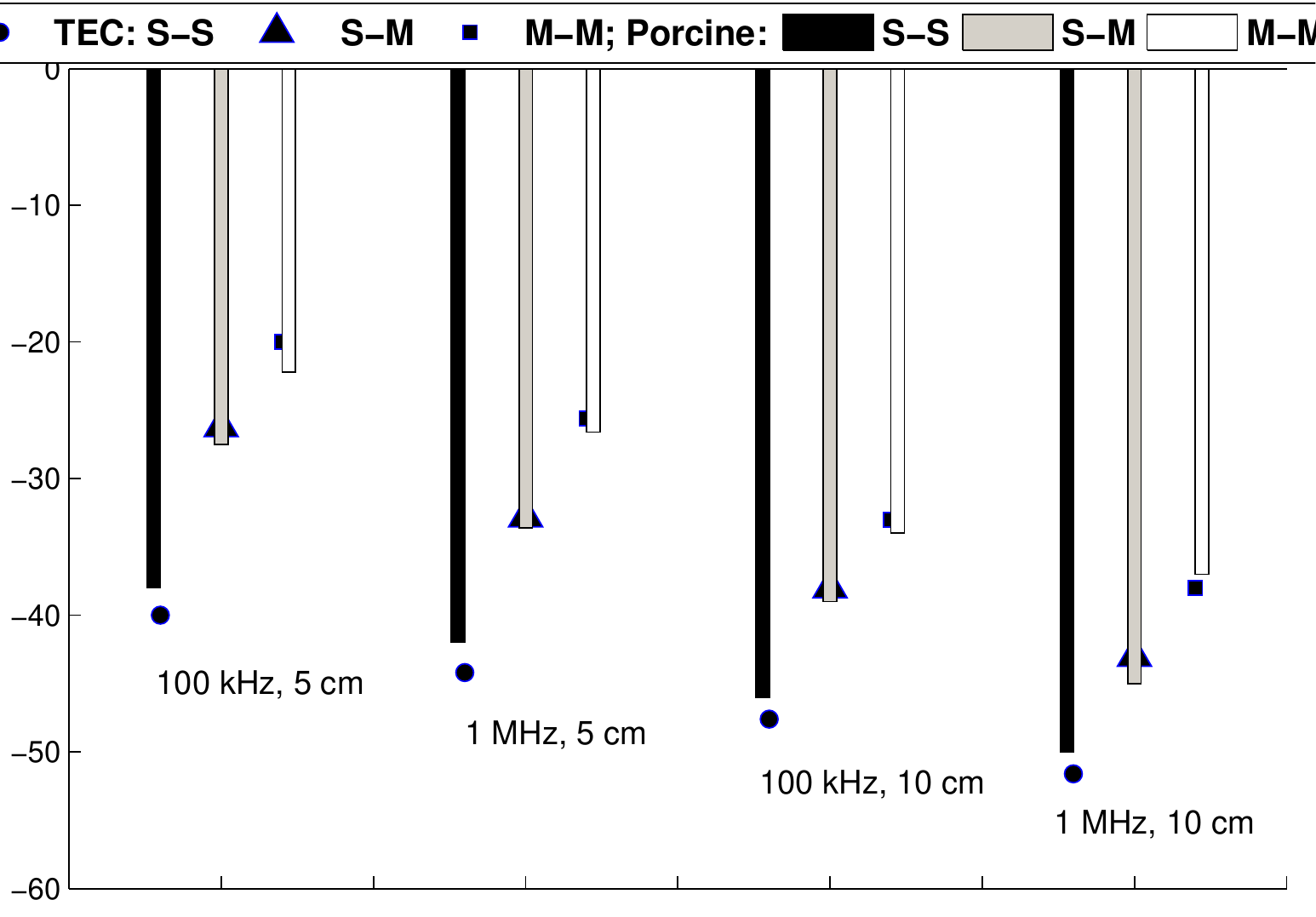}
 \caption{ \label{fig:pigplot} Porcine tissue experimental measurements Vs TEC model results}
 \end{figure}
\subsection{Discussion on Experimental Results}
The channel gain obtained using our analytical model, constructed without the bone layer, is compared against the real measurement made on the porcine tissue using the above given set-up and the average results obtained within $30\,\mathrm{min}$ and within $3\,\mathrm{hrs}$ of sacrifice are given in Fig.\ref{fig:pigplot}. The model gain along the M-M path out performs the S-S path by about $18\,\mathrm{dB}$ in $5\,\mathrm{cm}$ and $14\,\mathrm{dB}$ in $10\,\mathrm{cm}$ in both the test frequencies. The S-M and M-S path gains are close to each other as the muscle tissue is also exposed to the air, like the skin tissue, and there is no reflection from the bone tissue. Hence we consider them together in this study as S-M/M-S path. The empirical results are close to the TEC model results, which validates our approach.
\begin{table}[b]
\centering
\caption{\label{tab:tissue_change} Change in $G$ with tissue state and duration after excision}
\begin{tabular}{c|c|ccc}
\hline
Duration 	& State of& \multicolumn{3}{c}{ Deviation(dB)}\\ 
after excision & tissue & \multicolumn{3}{c}{from TEC model}\\
& & S-S	&S-M&	M-M\\
\hline
\textless 30  min&	Dry	&-1&	1&	2\\
& Moistened	&-4	&-1.5&	-0.5\\
\hline
2 to 3 hrs&	Dry	& 2&	7.5	&5.5\\
& Moistened	&-3	&5.5&	1\\
\hline
\end{tabular}
\end{table}

\tc{Albeit there are similarities between the human and porcine tissue, there are few differences that affect the accuracy of the TEC model. The porcine skin is relatively hairless and tightly attached to subcutaneous tissues. It is less vascular and also thicker. For instance, the stratum corneum of human skin is on average around $10\,\mathrm{\mu m}$ in thickness while that of porcine is $20\,\mathrm{\mu m}$. Similarly, the pH of porcine skin is $6-7$ and that of human skin is $5$. To add to this, the conductivity of muscle and fat varies from animal to animal by $\pm 0.1\ \mathrm{S/m}$ and $\pm 0.05\ \mathrm{S/m}$ in the range of frequency used \cite{prop1}. Also, the change in tissue properties over time caused by the variation in tissue hydration level and temperature \cite{pprop} as illustrated in Table.\ref{tab:tissue_change} contributes to measurement uncertainties as discussed below.}
 
\tc{When the tissue sample is freshly obtained (within $30\,\mathrm{min}$), the S-S path offered $1\,\mathrm{dB}$ more than the TEC gain with dry skin (refer Table.\ref{tab:tissue_change}). This is likely due to the abrasion on skin, caused by the shaving process that helped reducing the skin impedance. The impedance was further reduced when the locations of electrode attachment were moistened. However, the same measurements when observed after a couple of hours indicated a fall of $3\,\mathrm{dB}$ from the initial gain. Moistening the skin helped recovering $5\,\mathrm{dB}$ of gain compared with the dry tissue state. We obtained the average value of these measurements in each path and plotted them in Fig.\ref{fig:pigplot}. There is a difference of $3\  \mathrm{dB}$ between the analytical model and empirical results, which is likely contributed by the above mentioned uncertainties, the reasons highlighted in Section.\ref{Results}, and due to the structural damage caused by excision.}
 
\section{Conclusions}\label{Conclusion}
IBNs will lead to diverse health care applications that would benefit at health risk populations and patients at remote locations when the presence of a human caregiver or trained medical professional is not always possible. The ability to sense physiological changes within the body and take proactive monitoring steps will increase human longevity at reduced health care costs. As a first step towards the galvanic coupled IBN described in this paper, we derived, verified and validated the equivalent electrical circuit model for human tissues in characterizing the physical layer. We conducted extensive studies regarding the gain and phase-change in the transmitted signal under varying operating frequencies, tissue dimensions, sensor placements, electrode separation distances and dimensions, among others, to comprehensively characterize the body channel, while respecting permissible safe current limits. 

We found that a maximum of $30\,\mathrm{dB}$ in channel gain could result from variation in tissue properties from person to person. We identified the optimal frequency to lie between $100\,\mathrm{kHz}$ to $1\,\mathrm{MHz}$ for both on skin and in muscle paths, and determined that placing both the sender and receiver sensors within the muscle offered better channel propagation characteristics, as opposed to on the skin. We will investigate future topics in wireless communication, including derivation of achievable capacity and optimal modulation schemes, along with higher layer protocol design using the channel models derived in this paper. 
\section*{Acknowledgements}
\label{ACK}
This material is based on work supported by the U.S. National Science Foundation under Grant No. CNS-1136027 and CNS-1453384. The authors are grateful for the helpful discussions and inputs provided by Deniz Erdogmus from Northeastern University and Taskin Padir from Worcester Polytechnic Institute. 

\vspace{-1cm}
%\bibliography{IEEEabrv,p1}
\begin{IEEEbiography}[{\includegraphics[width=25mm,height=28mm,clip,keepaspectratio]{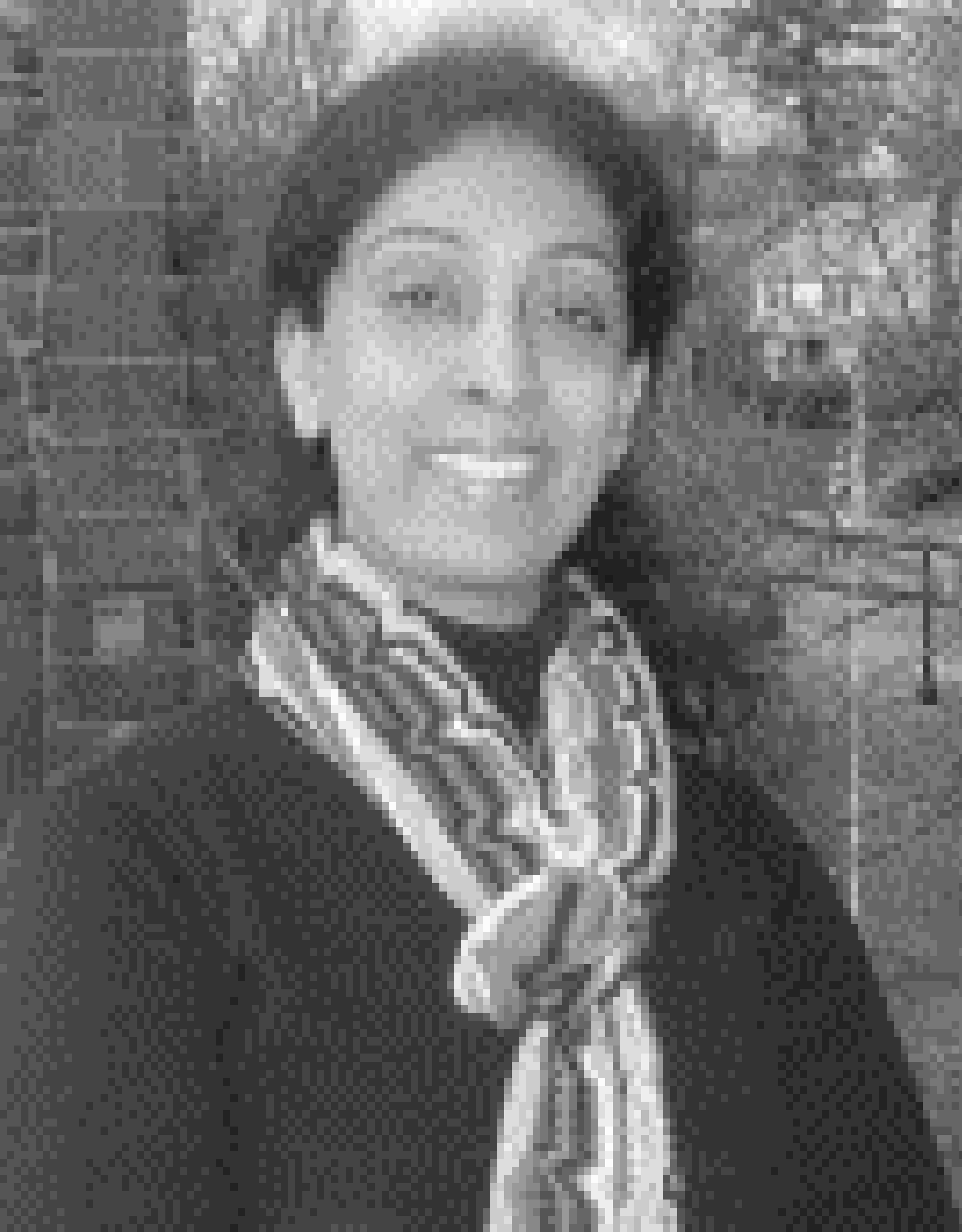}}] {Meenupriya Swaminathan} is a research assistant with the department of electrical and computer engineering, Northeastern University, USA, where she is also currently pursuing the Ph.D. degree on galvanic coupled intra-body wireless communication. She received the B.E. degree in electronics and communication engineering from Bharathidasan University, India, and the M.E. degree in computer engineering from Anna University, India. Her research interests include tissue modeling, intra-body communication and energy awareness in wireless network protocols.
\end{IEEEbiography}
\vspace{-1cm}
\begin{IEEEbiography}[{\includegraphics[width=25mm,height=28mm,clip,keepaspectratio]{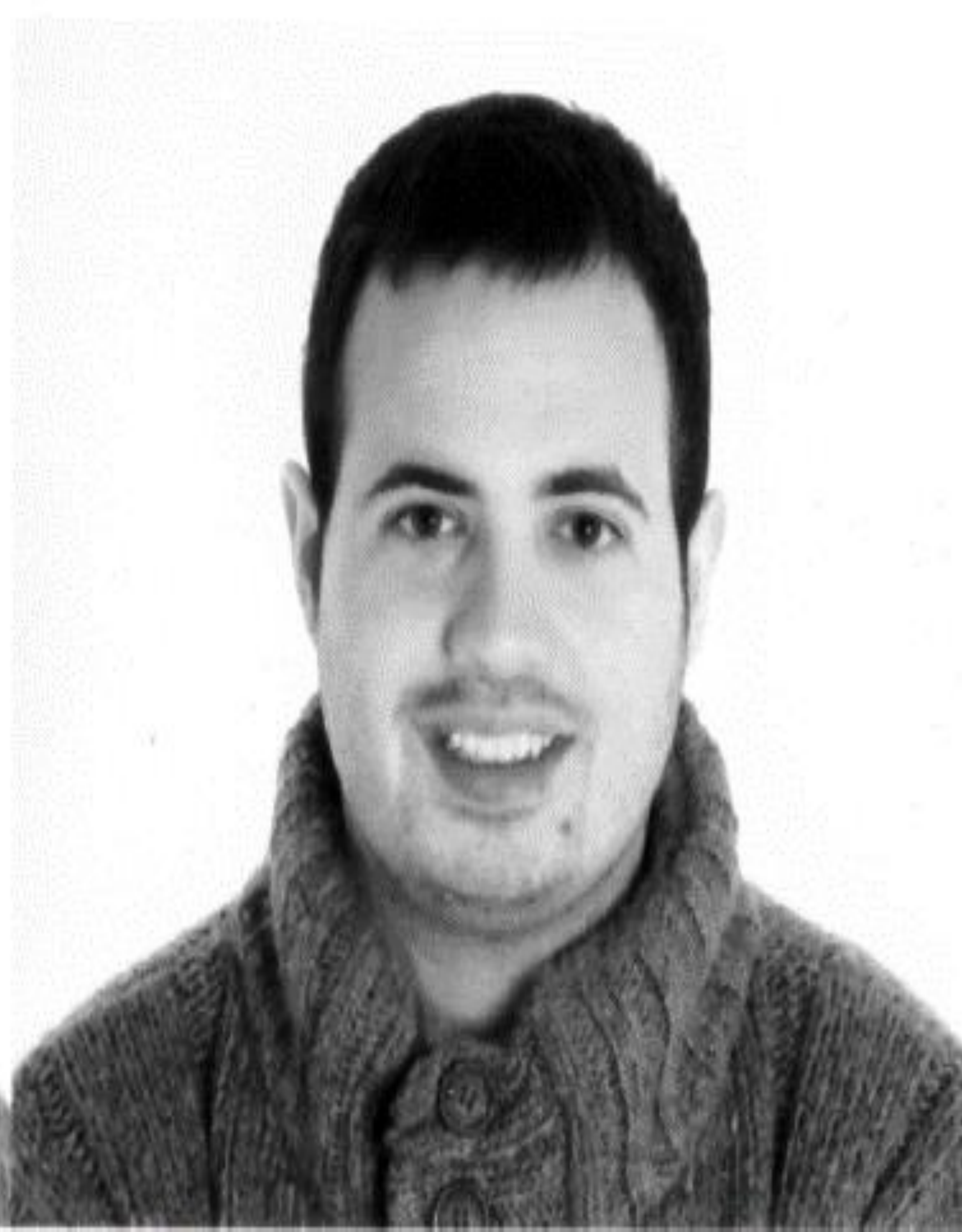}}] 
{Ferran Simon Cabrera} received his B.S. degree in telecomunications engineering and M.S. in electronics engineering from Universitat Politecnica de Catalunya, Bacerlona, Spain, in 2011 and 2014, respectively. 
He was a visitor student in the National Institute of Informatics, Tokyo, Japan, in 2011 and in Northeastern University, Boston, USA, in 2013 developing this project.  He is currently a developer at an internet security company in Barcelona, Spain. His research interests include renewable energy, optical communications and microelectronic design.
\end{IEEEbiography}
\vspace{-1cm}
\begin{IEEEbiography}[{\includegraphics[width=25mm,height=28mm,clip,keepaspectratio]{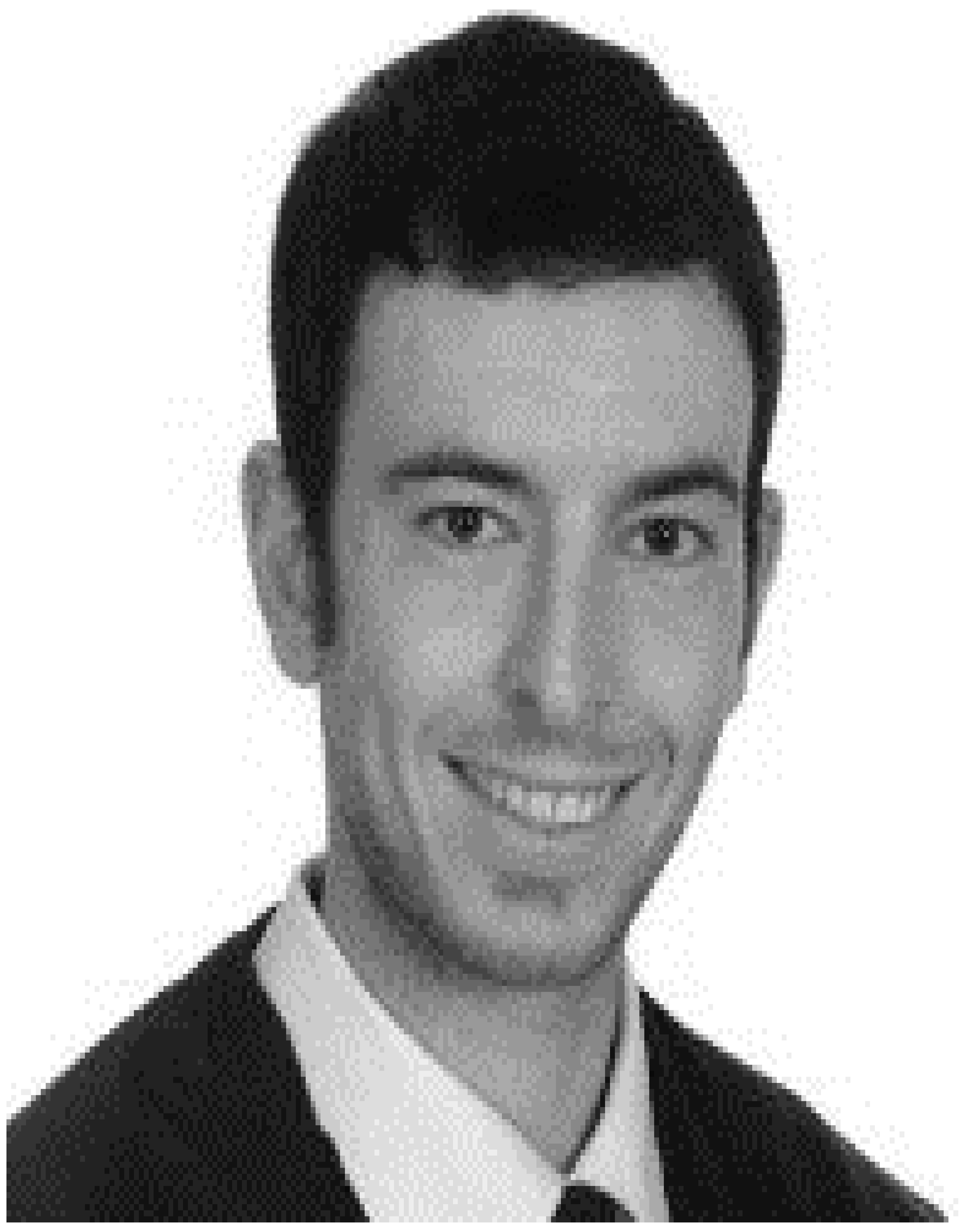}}] {Joan S. Pujol} received the B.S. degree in telecommunications engineering from Universitat Politècnica de Catalunya, Barcelona, in 2014, where he is currently pursuing M.S. degree in telecommunications engineering. His research interests include wireless communications, intra-body communication and signal processing.
\end{IEEEbiography}
\vspace{-1cm}
\begin{IEEEbiography}[{\includegraphics[width=25mm,height=28mm,clip,keepaspectratio]{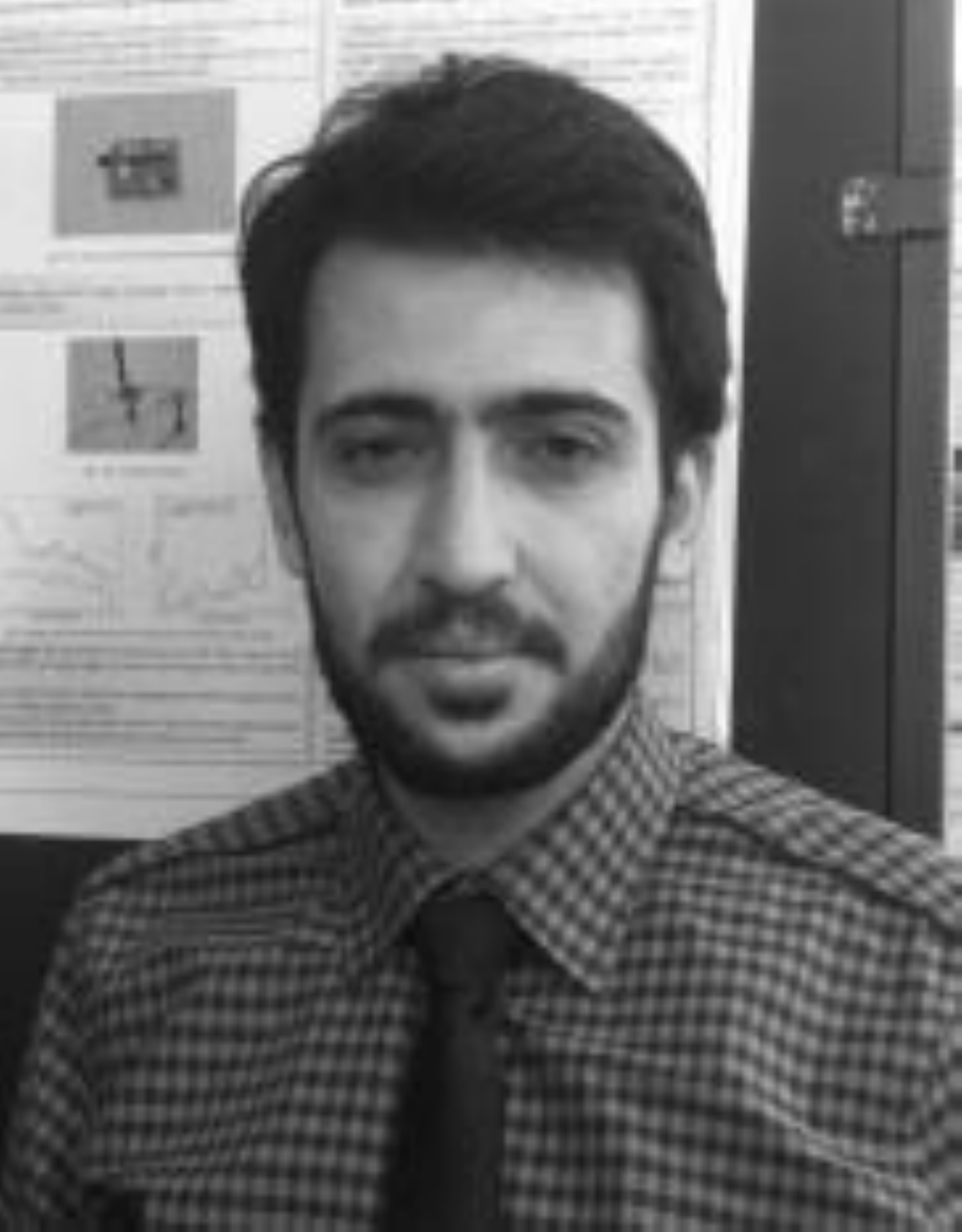}}]
{Ufuk Muncuk} is a research assistant with the department of electrical and computer engineering, Northeastern University, Boston, USA, where he received his M.S. degree in computer science and is currently pursuing the Ph.D. degree on wireless energy transfer system design. He received the M.S degree and the B.S. degree in electrical and electronics engineering from Erciyes Üniversitesi, Turkey and Fırat Üniversitesi, Turkey, respectively. His research interests include cognitive radio systems and RF energy harvesting systems. 
\end{IEEEbiography}
\vspace{-1cm}
\begin{IEEEbiography}[{\includegraphics[width=25mm,height=28mm,clip,keepaspectratio]{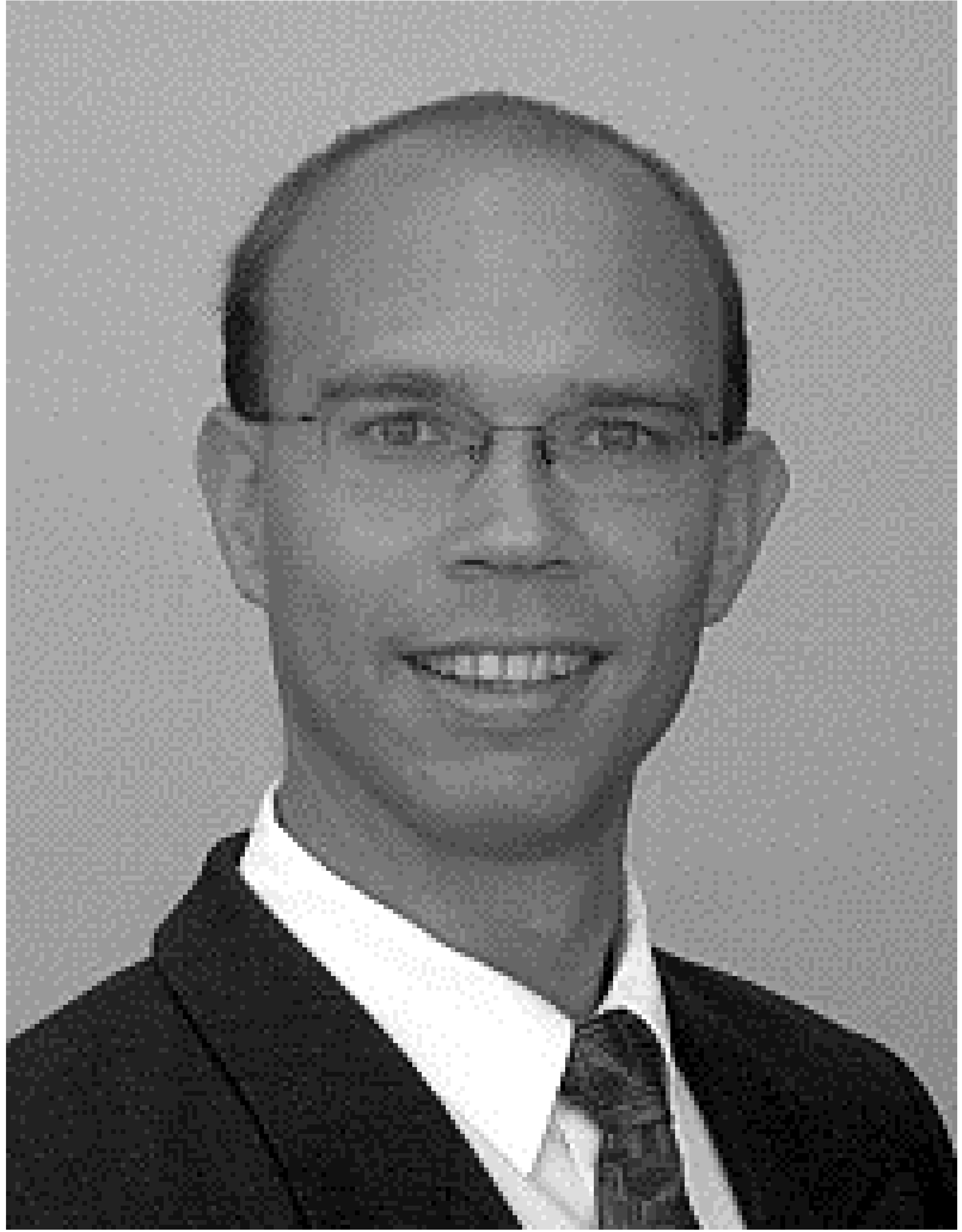}}]
{Gunar Schirner} is an assistant professor in the Department
of Electrical and Computer Engineering at Northeastern
University. His research interests include embedded system
modeling, system-level design, and the synthesis of embedded
software. Schirner received a PhD in electrical and
computer engineering from the University of California,
Irvine. He is a member of IEEE. 
\end{IEEEbiography}
\vspace{-1cm}
\begin{IEEEbiography}[{\includegraphics[width=25mm,height=28mm,clip,keepaspectratio]{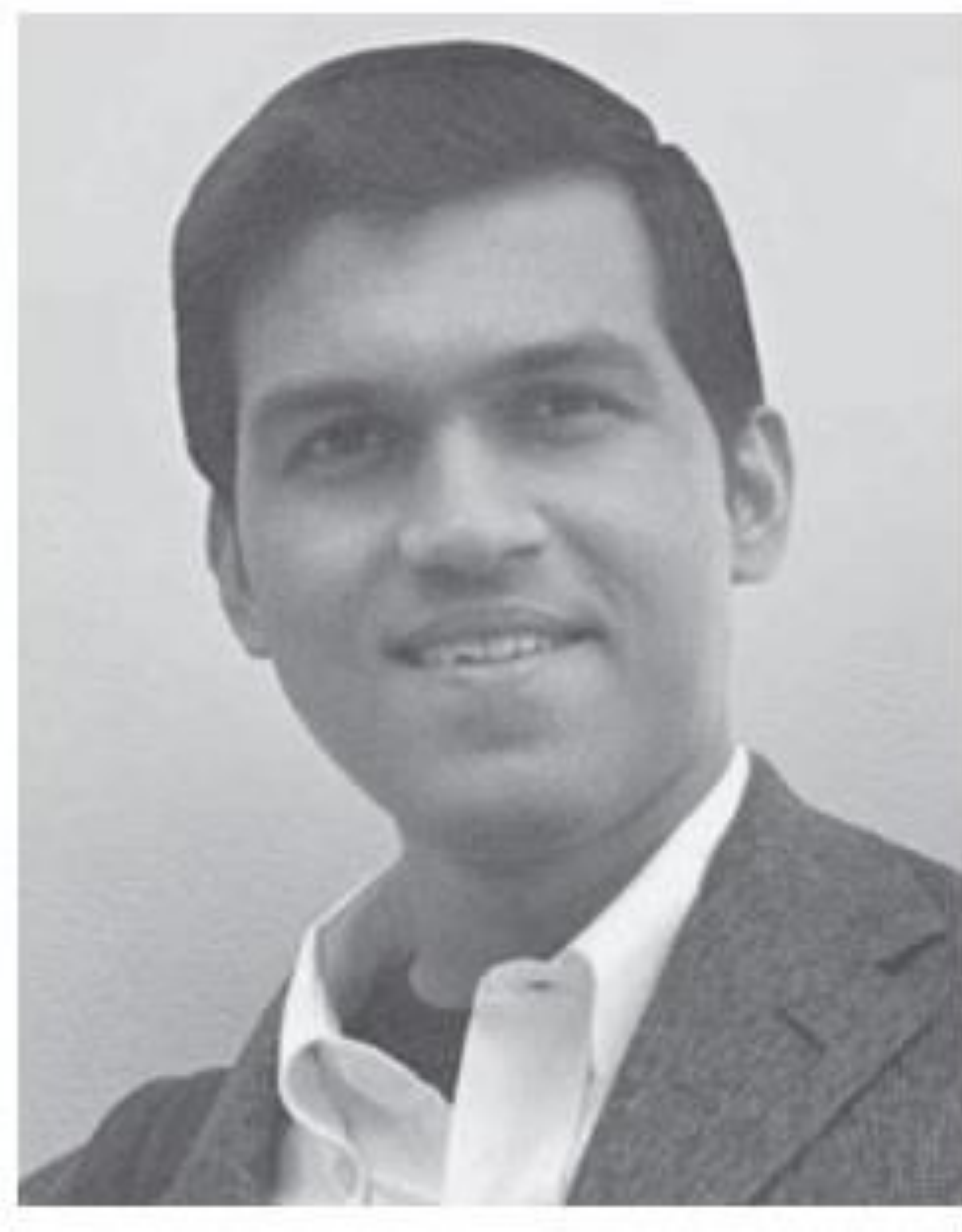}}]
{Kaushik R Chowdhury} [M'09] received his B.E. degree in
electronics engineering with distinction from VJTI, Mumbai
University, India, in 2003, his M.S. degree in computer science from the University of Cincinnati, Ohio, in 2006, and his Ph.D. degree from the Georgia Institute of Technology, Atlanta, in 2009. His M.S. thesis was given the outstanding thesis award jointly by the Electrical and Computer Engineering and Computer Science Departments at the University of Cincinnati. He is an assistant professor in the Electrical and Computer Engineering Department at Northeastern University. He currently serves on the editorial board of the Elsevier Ad Hoc Networks and Elsevier Computer Communications journals. His expertise and research
interests lie in wireless cognitive radio ad hoc networks,
energy harvesting, and intra-body communication. He is
the recipient of multiple best paper awards at IEEE ICC.
\end{IEEEbiography}
\end{document}